\documentclass[12pt]{article}
\usepackage{amssymb,amsmath,latexsym}
\usepackage[english]{babel}%
\usepackage{multirow}% http://ctan.org/pkg/multirow
\usepackage{hhline}
\usepackage[utf8x]{inputenc} 
\usepackage{booktabs} % To thicken table lines
\usepackage{footmisc}
\usepackage{float}
\usepackage{makeidx} 
\usepackage{footnote}
\usepackage{graphicx} 
\usepackage{hyperref}
\usepackage{cleveref} % this package must be used after hyperref. 
\usepackage{epstopdf}
\usepackage{subfigure}
\usepackage{setspace}
\usepackage{pdflscape}
\usepackage{csquotes}
\usepackage{url}

\usepackage[compatible]{nomencl}
\usepackage{tikz}
\usetikzlibrary{shapes.geometric, arrows}

\begin{document}
\title{Determination of water permeability for cementitious materials with minimized batch effect}
\author{
Zhidong Zhang, George Scherer \\
\small
\textit{Department of Civil and Environmental Engineering, Princeton University, Princeton, NJ 08544, USA}
	}
\date{\vspace{-5ex}}
\maketitle

\begin{abstract}
Values of water permeability for cementitious materials reported in the literature show a large scatter. This is partially attributed to the fact that materials used in different studies are different. To eliminate the effects of cements, specimen preparation, curing conditions and other batch effects, this study employs a long cylindrical cement paste to prepare all specimens for a variety of permeability determination methods, such as beam bending, sorptivity, Katz–Thompson and Kozeny-Carman equations. Permeabilities determined by these methods are then used in a moisture transport model. Compared with the measured mass loss curves, we found that permeability determined by the beam bending method provides much closer results to the measured ones than other methods. The difference results from that the saturated specimen is used in the beam bending method while specimens in other methods are dried (or rewetted). As already shown in the literature, the microstructure of the dried or rewetted specimens is altered and different to the original microstructure of the water saturated specimens.  \ldots 
\end{abstract}

\section{Introduction} 
\label{section:introduction} 
% Importance of liquid water transport
The durability of concrete structures is always closely related to the moisture transport properties in cementitious materials. The liquid uptake when concrete is in contact with liquid water (e.g., groundwater, rain) can induce the penetration of aggressive agents (e.g., chloride ions) through the concrete cover. A common natural condition - drying/wetting cycles - can increase the rate of chloride ingress compared to the saturated condition~\cite{Hong1999}. Carbon dioxide may transport within the gaseous phase in concrete and decrease pH of the pore solution. All these processes are able to result in corrosion to the rebars and deterioration of concrete. For this reason, moisture transport becomes a crucial theme when evaluating the durability. 

% Modelling of moisture transport in concrete 
Moisture transport in partially saturated porous media such as cementitious materials is mainly governed by the transport of three phases: liquid water, water vapour and dry air. The previous studies have shown that dry air has very low contribution to the mass of moisture transport and only causes fluctuating air pressure in the material~\cite{Mainguy2001,Zhang2012}. This conclusion was also drawn by the asymptotic analysis performed by Coussy and Thi\'ery~\cite{Coussy2010,Coussy2009}. In addition, considering that the liquid phase remains incompressible and total gas pressure is constant, the mass balance equations of moisture transport can be represented by a simplified equation, including only liquid water and vapour~\cite{VBB2007b,Zhang2015}. Mainguy et al. \cite{Mainguy2001} further simplified the model for specific conditions, by considering only liquid water and neglecting the vapour diffusion. They found that such a model can give results for simulating drying mass loss curves very similar to the multiphase model. Hence, the transport coefficient - water permeability that governs the liquid water transport becomes extremely important. For a given material, the ideal situation to perform moisture transport simulations is that $K_l$ is experimentally determined and then used in the moisture transport model. 

% What is intrinsic/saturated permeability
% One of the moist important parameters for moisture transport is the liquid permeability $K_l$ (also known as hydraulic conductivity) which is used in either Darcy's law for saturated transport or extended Darcy's equation by Buckingham~\cite{Buckingham1907} for describing water movement in unsaturated non-swelling porous media. 
%For unsaturated solids, the mass flux $J_l$ ($\mathrm{kg\cdot m^{-2}\cdot s^{-1}} $) is written as: 
%\begin{eqnarray}\label{eq:Darcy1} 
%J_l = - \rho_l k_{rl} \dfrac{K}{\eta} \nabla P_l 
%\end{eqnarray}
%where $ \rho_l $ ($ \mathrm{kg\cdot m^{-3}} $) is the density of liquid water, $\eta$ (Pa $\cdot$ s) represents the dynamic viscosity of liquid water, $P_l$ (Pa) is liquid pressure and $k_{rl}$ is the relative liquid permeability which is generally treated as a function of moisture content or degree of saturation.  
%

In most literature, water permeability $K_l$ with a unit of m$^2$ is considered as an intrinsic property of the porous material, meaning that $K_l$ only depends on the microstructure and should be valid for other fluids. That is why in some papers $K_l$ is named as \enquote{intrinsic} permeability (e.g.,~\cite{Mainguy2001,Zhang2015,Zhang2016}). Nevertheless, \enquote{intrinsic} permeabilities measured by gases (oxygen and nitrogen~\cite{Zhou2017}) and solvents (methanol~\cite{Loosveldt2002} and isopropanol~\cite{Hearn1996,Zhou2016indirect}) are often found to be greater than $K_l$. Reasons for this will be discussed later. To avoid the confusion, in this paper, the terminology of water permeability instead of intrinsic permeability is used hereafter. 
%Permeability is the key parameter for modelling of moisture transport. 
%Results reported in the literature scatter over a large range for the similar materials. 
%First, the moisture transport model will be introduced. 

% Review of measurement methods  
For cementitious materials, the determination of water permeability is an active research field reported in a voluminous literature. 
%As we have shown, there are many permeability test procedures although preferred, recommended and standard methods have not yet clearly emerged. That permeability is so widely investigated reflects the global interest in the durability of concrete which is rightly seen as being intimately associated with water migration and water-mediated chemical reactions. Most water transport in cement-based materials occurs by means of unsaturated flow, for which it is necessary but not sufficient to know the saturated permeability.
The permeability can be determined either directly by experiments or indirectly by theoretical models based on other measured data. 
Conventional methods to measure $ K_l $ are classified as flow-through techniques as they measure the flux under steady state condition for fully saturated specimens, with the geometry of either truncated cones~\cite{Nyame1981} or cylinders/disks~\cite{Hearn1991,ElDieb1995,Bagel1997,Ye2005,Phung2013determination,Kameche2014,Zhou2016indirect}. During measurements, the side of the specimen must be sealed and liquid water is injected from one face by applying extra pressure so that the outflow can be only observed on the opposite face. When the flow in the porous body reaches steady state, the flow rate is then used to calculate $ K_l $ according to Darcy's law.  %However, the disadvantages of the conventional methods are obvious. They require very high extra pressure. For example, in high performance concretes with low water-to-cement ratios and the fine pore structures, moisture transport is much slower than the traditional ordinary cementitious materials and the amount of liquid water flowing out of the specimen is very low. For such low-permeability porous media, one must increase the applied pressure; meanwhile, this also incredibly increases the risk of physical damage of the pore structure. Besides, under high pressure, the leakage is unpredictable through the sealant and the rough surface of specimen. In addition, the essentially saturated state which requires prior vacuum is very difficult to reach, especially for low water-to-cement ratio material. % More recently, an improved device using a controlled oedometer seems to be able to avoid the leakage problem, but it still needs a high extra pressure~\cite{Picandet2011}.  
These methods are not difficult to do, although they may take a long time to reach steady state flow (e.g., several weeks) for low permeable materials. To reduce the measurement time, it was suggested to increase the applied pressure~\cite{Kameche2014}, but this may risk altering the structure of materials and increase the water leak at the interface between the specimen and the pressure cell. Instead of applying continuous constant pressure, pressure relaxation methods involve increasing pressure on one side and observing the decrease of pressure due to liquid being pushed to another side (e.g.,~\cite{Brace1968}). These methods are rapid but they still need high pressure and thus have the same problems as the other conventional flow-through methods. 

Recent studies used hollow cylinders~\cite{Jones2009,Rose2017,Amriou2017} which measured radial flow of water under applied pressure.
The main advantage of this kind of method is that the total area which allows fluids flowing through is much larger than the disc specimens; therefore, the measurements showed higher accuracy and repeatability~\cite{Jones2009,Amriou2017}, whilst the large area may have higher chances to face the effect of heterogeneity, which means that any cracks or area having greater water flow can significantly change the results. % These methods require good homogeneity for the specimen. 
Another method is the dynamic pressurization (DP) which keeps the specimen in a sealed vessel and suddenly increasing or decreasing pressure~\cite{Scherer2006,Grasley2007}. By alternatively pressuring and depressuring, the effect of air voids in the unsaturated specimen can be gradually removed~\cite{Grasley2007}.

Indirect methods, requiring other data that can be used to calculate $K_l$, are referred as poromechanical (dynamic pressurization) techniques which monitor the time-dependent deformation of a specimen related to fluid flow in the pore network induced by externally applied stress or temperature change. The beam bending (BB) method is one such rapid indirect methods~\cite{Scherer1992,Scherer2000a,Vichit2002,Vichit2003} which was originally developed for soft gels and later was applied to cementitious materials. This method is based on the principle of exerting a certain strain to a long and slender specimen to obtain a relaxation curve, which is considered including both hydrodynamic and viscoelastic effects. By fitting this relaxation curve, $ K_l $ can be determined. This method has very clear requirements for the geometry of the specimen which makes it less applicable for concretes due to needing inconveniently large specimens to obtain a representative volume including aggregates. 
A method so-called thermopermeametry (TPA) was also introduced to cementitious materials on the basis of research about gels~\cite{Scherer1991,Scherer1994,Scherer2000b,Ai2001}. Given that the thermal expansion of liquid is always much greater than that of the solid phase, this method can determine $ K_l $ by measuring the rate of strain relaxation as liquid water flows out of the material after a temperature change.  %If repeating the pressure cycles, the residual air in the material can be slowly dissipated. However, under high external hydrostatic pressure mechanical deformation may occur over a long period and the solid skeleton may be damaged during depressurization. % The latest literature revealed that the NMR technique is a good way to analyse liquid permeability~\cite{Zamani2014}. Unfortunately, there is no data available for liquid permeability in the literature.  

% the general problem for measurements 
Measurements of $K_l$ are very sensitive to saturation conditions since the fully saturated condition is not easy to achieve. The presence of air voids or entrapped air in non-fully saturated materials may have a great influence on the measured results of the poromechanical methods and long delays in reaching equilibrium in conventional methods~\cite{Scherer2008}. To ensure the fully saturated condition, various approaches were used in the literature, such as curing the specimen in water/limewater~\cite{Amriou2017,Zhou2016indirect}, vacuum saturation~\cite{Nguyen2009} and pressurizing saturation~\cite{Vichit2002,Vichit2003}. 
The time needed to fully saturate a porous body increases with the square of its smallest dimension. For the direct methods, the thickness of a disc specimen ranges from 25 to 70 mm (see the review in~\cite{Kameche2014}) depending the size of aggregates as El-Dieb and Hooton~\cite{ElDieb1995} suggested that the specimen thickness should be 3 times as large as the aggregate size and a recent study~\cite{Wu2015} even reported that the specimen needs to be about 10 times as thick as the aggregates; therefore, a specimen may be extremely difficult to saturate. By contrast, the specimen in BB measurements is much easier to saturate since the method is limited to paste and mortar, so the diameter of the cylinder can be smaller than the concrete specimens used in the conventional methods. 

% review of theoretical models 
In addition to experimental measurements, water permeability can be inferred by using information related to the microstructure, such as pore size distribution (PSD), porosity, tortuosity, connectivity, etc. A relationship was first proposed by Kozeny in 1927~\cite{Kozeny1927} and later modified by Carman~\cite{Carman1937,Carman1956} which is commonly known as the Kozeny–Carman (KC) equation. This equation was developed after considering a porous material as an assembly of capillary tubes for which the Navier–Stokes equation can be applied. It yielded $K_l$ as a function of the porosity, the specific surface, and the shape and tortuosity of channels. It has been found that the KC equation is approximately valid for sands but not for clays~\cite{Lambe1969}. 
Wong et al.~\cite{Wong2006} adopted a modified KC equation incorporating tortuosity and constrictivity to predict the oxygen permeability for concrete and they concluded that this equation overestimated the permeability by about one order of magnitude. 

Based on mercury intrusion porosimetry (MIP) data for assessment of the percolation radius of the microstructure and resistivity measurements to determine the formation factor (inversely proportional to the product of porosity and tortuosity), the Katz-Thompson model~\cite{KT1986} (KTI model) has proven useful for estimating the permeability of sedimentary rocks. Another version of this model avoids the use of resistivity measurements by assessing the formation factor only using mercury intrusion (KTII model)~\cite{KT1987Prediction}. %The Katz-Thompson method was used for cementitious materials~\cite{VBB1999,Mainguy2001,VBB2011b}.
Baroghel-Bouny et al~\cite{VBB2011b} showed that the KTI relationship slightly overestimates the permeability for concretes and mortars. The investigations of KTII performed by El-Dieb and Hooton~\cite{ElDieb1994} illustrated that this model leads to a more pronounced overestimation; the correlation can even be quite low for cementitious materials. Recent study by Zhou et al.~\cite{Zhou2017} showed that KTII can provide similar results to measure nitrogen gas permeability with Klinkenberg correction, but it is about 2-4 orders of magnitudes higher than water permeability. 

Other methods based on theoretical models are also reported in the literature. For instance, a practical method is to use measured sorptivity to assess $K_l$~\cite{Zhou2016indirect} because sorptivity measurements are much easier to perform than the above-mentioned permeability measurements. %Like the KT equations, this method largely relies on if the theory is applicable to cementitious materials. %In~\cite{Zhou2016},  on the models. 
The first author of this paper introduced two methods to indirectly determine water permeability~\cite{Zhang2016}. One is called \enquote{inverse analysis} that utilizes a numerical moisture transport model to back calculate $K_l$ based on the measured drying mass loss curve. The other one employs the measured diffusivity curve to fit $K_l$ by a general expression including both liquid transport and vapour diffusion. 

% Problems, the present study only focuses on the effect of specimen
As stated above, various methods using different specimen geometry or theoretical models in the literature to determine $K_l$ show that the reported permeability values have great dispersion, ranging from 10$^{-22}$ to 10$^{-17}$ m$^2$ for materials with the same water-to-cement ratio (see review in~\cite{Zhang2016}). % Even the gas permeability, which is easier to measure than water permeability, can dramatically change with specimens (even coming from the same batch of concrete \cite{AitMokhtar201321}). 
Even though researchers are apt to compare these values from different studies, $K_l$ determined by various methods is not directly comparable, because: \\
\indent (1) the cements are different (the chemical composition of cements from different cement plants vary), \\
\indent (2) specimen preparation procedures are different (although the same preparation procedure is claimed in different studies, the variations of experimentalists, equipment, environment, etc, are unavoidable),  \\
\indent (3) and specimens curing methods are different. 

To eliminate these artificial effects, the present study employs a slender specimen for various methods to determine water permeability $K_l$. This long cylinder was initially prepared for the BB method. After the BB measurements, it was cut into several short cylinders and many slices. Some of short cylinders were used in sorptivity measurements and the others were used in drying experiments which provided the calibration data for a moisture transport model. Slices were used to measure the desorption isotherm which serves as the input data for the numerical model. Meanwhile, a small part of this slender cylinder was crushed to prepare specimens for MIP, nitrogen adsorption (NAD) and thermogravimetric analysis (TGA) tests. % The water permeability $K_l$ reported in the literature were rarely used to verify a moisture transport model as the model need more experimental data to be verified. 

The structure of this paper is given as follows. Firstly, a moisture transport model with $K_l$ as the only undetermined parameter will be introduced, and then methods used to determine $K_l$ will be briefly described. Experiments are carried out to obtain data for permeability determination methods and the moisture transport model. Finally, results from these methods will be compared and discussions on these methods will be presented. 

% To take into consideration the potential variability property of concrete, the experiments were devised such that all results were from the same mixes. In addition, the materials for the concrete mixes were all from the same batch to further reduce potential discrepancies due to materials properties. 

% Scope of the present study 

%%%%%%%%%%%%%%%%%%%%
% NEW SECTION
%%%%%%%%%%%%%%%%%%%%
\section{Moisture transport model}
\label{section:Modelling}
\subsection{Governing equations}
Even though the multiphase model can be simplified as a single-phase model with liquid water~\cite{Mainguy2001}, the authors also pointed out that this simplified approach is only suitable for the case of drying of low permeability materials with the initial condition close to saturation and exposed to a high relative humidity (RH) boundary condition. In that situation, liquid water governs mass transport~\cite{Zhang2012,Zhang2017DT}, while in a lower RH range vapour diffusion is non-negligible. The model selected here to simulate moisture transport in cementitious materials is the semi-simplified version of the multiphase model - including the transport of both liquid water and water vapour - which considers that gas pressure is constant and the liquid phase remains incompressible. The governing equation for the mass balance is written as~\cite{VBB2007b,Zhang2015,Zhang2016} 

\begin{equation} \label{eq:mass_balance}
\rho_l \dfrac{\partial S}{\partial t}  = \frac{1}{\phi} \ \mathrm{div} \ ({J}_l +  {J}_v )
\end{equation}
where $S$ is the degree of saturation, $ \rho_l $ ($ \mathrm{kg\cdot m^{-3}} $) is the density of liquid water, $ \phi $ is the porosity of the porous material, and ${J}_l$ and $ {J}_v $ ($\mathrm{kg}\cdot \mathrm{m}^{-2}\cdot \mathrm{s}^{-1}$) are the fluxes of liquid water and water vapor, respectively.

Liquid water transport consists of both contributions of capillary viscous movement of free water and the transport of physically-adsorbed water molecules in a single Darcy relation in which capillary pressure and relative permeability are derived from the sorption curves over the whole range of RH through the liquid water saturation $S$ (-). Moreover, the viscosity of pure water is used for both free water and bound water. The driving force for liquid transport is the gradient of liquid pressure $ P_l $ ($ \mathrm{Pa} $) \cite{Mainguy2001, VBB2007b}:

\begin{equation} \label{eq:liquid_tansport}
{J}_l = -\rho_l \dfrac{K_l k_{rl} (S)}{\eta} \mathbf{grad}\, P_l
\end{equation}
where $\eta$ (Pa $\cdot$ s) represents the dynamic viscosity of liquid water, and $k_{rl}$ (-) is the relative liquid permeability which is generally treated as a function of $S$.  

The flux of vapor $ {J}_v $ is described as a diffusion-like process with the vapor density $\rho_v$ ($ \mathrm{kg\cdot m^{-3}} $) as the main variable~\cite{Mainguy2001, Zhang2012}. %decomposed in two terms: the first component corresponds to the diffusion contribution and the second to the advective one~\cite{Mainguy2001}. If assuming that the gas pressure in the material remains almost constant (equal to the atmospheric pressure), it means that the advective contribution can be neglected~\cite{Mainguy2001, Zhang2012}. Thus, $ \mathbf{w}_v $ is written as:

%  If assuming that the gas pressure in the materials equal to the atmospheric pressure ($ P_g = P_{atm} $), it means that there is no potential for the advection of gas-phase, so the diffusion of gas-phase is governed by the gradient of gas concentration~\cite{Mainguy2001, Zhang2012}. Because water vapor predominates the mass of gas-phase, the diffusion of gas-phase corresponds to a pure water vapor diffusion. Hence, Fick's first law is sufficient to describe this process~\cite{Mainguy2001, VBB2007b}.   
% 
%\begin{equation} \label{eq:vapor_diffusion}
%\mathrm{w}_v = -D_{v0} \rho_g f(S, \phi) \mathrm{grad} \dfrac{\rho_v}{\rho_g} 
%\end{equation}
%
%In Eq.~(\ref{eq:vapor_diffusion}), the driving force of diffusion is the gradient of $ \rho_v$ since $ P_g= $ constant. %  $\dfrac{\rho_v}{\rho_g} $

\begin{equation} \label{eq:vapor_diffusion}
{J}_v = -D_{v0} f(S, \phi) \ \mathbf{grad} \ \rho_v
\end{equation}
where $ D_{v0} $ ($ \mathrm{m^{2} \cdot s^{-1}} $) is the free vapor diffusion coefficient in the air. The parameter $ f (S, \phi) $ represents the resistance factor for gaseous diffusion and is related to the connectivity and tortuosity of the pore network. 

In this moisture transport model, the thermodynamic equilibrium between the liquid and vapor is assumed. The equilibrium state is governed by Kelvin's law which is written in the following form:

\begin{equation}\label{eq:kelvin}
\displaystyle P_c = - \frac{ \rho_lRT }{ M_v } \ln \mathrm{RH} 
\end{equation}
where $ R=8.314 \text{J}\cdot \text{K}^{-1} \cdot \text{mol}^{-1} $ is the gas constant, $ T $ ($ \mathrm{K} $) is the absolute temperature and $ M_v $ ($ \mathrm{kg} \cdot \text{mol}^{-1} $) is the molar mass of water molecule. Capillary pressure in the macroscopic scale is defined as the difference between gas pressure the liquid phase pressure ($P_c = P_g - P_l$). The relationship of $P_c$ as a function of $S$ is known as the capillary pressure curve. For cementitious materials, this curve is indirectly measured by means of sorption experiments performed at constant temperature (so-called water vapor sorption isotherms) \cite{VBB2007a}. Various equations can be found in the literature to describe sorption isotherms~\cite{Zhang2014}. One well-known equation was proposed by van Genuchten (VG equation)~\cite{VG1980},

\begin{equation} \label{eq:VG2}
P_c (S) = \alpha \left( S^{-1/m} -1 \right)^{1-m}
\end{equation} 
where $ \alpha $ (Pa) and $ m $ are two fitting parameters. 

The flux boundary condition (also known as convective condition~\cite{Pont2011}) is used to account for an imperfect moisture transport between the environment and the surface of the material. The expression is given as~\cite{Nguyen2009} 

\begin{equation} \label{eq:boundary}
q = \left({J}_l + {J}_v\right)_{x=0} =  \phi S_0 E (P_v^0 - P_v^{e})
\end{equation}

This boundary condition includes a material property (porosity $ \phi $), a parameter related to the environment (external vapor pressure $ P_v^{e} $), the moisture state within the material near the surface ($ P_v^0 $ and its related liquid-water saturation $ S_0 $) and the interaction between the ambient environment and the material (through the emissivity $ E $). The term $ \phi S_0 $ takes into account the reduction of wet surface when exposed to the environment. The emissivity $ E $ ($ \mathrm{kg \cdot m^{-2} \cdot s^{-1} \cdot Pa^{-1}}$) has been assessed from experiments which proposed a value around $ 2.58 \times 10^{-8}\, \mathrm{kg\cdot m^{-2}\cdot s^{-1}\cdot Pa^{-1}}$~\cite{Azenha2007,Nguyen2009} for a laboratory environment where RH and temperature are maintained constant (atmospheric pressure, RH = $ 50\pm 5\% $ and $ T = 293\, \mathrm{K} $). %Since the present study only considers conditions without wind effects (the specimens are submitted to drying or wetting conditions in a desiccator), it is reasonable to assume that $f(v)=1$. Moreover note that the same value of $E$ is fixed for both drying and wetting modelling. 

Putting Eqs.~(\ref{eq:liquid_tansport}) - (\ref{eq:VG2}) back into the mass balance equation, Eq.~(\ref{eq:mass_balance}), and combining the initial and boundary conditions (see Eq.~\ref{eq:boundary}), the moisture transport problem can be solved. %Expressions for $\rho_v$, $P_v$ 
%It also should be noticed that Eq.~(\ref{eq:VG2}) is used for the main sorption curve. In the case of modelling with hysteretic effects, a hysteresis model is needed to calculate a new $ P_c - S $ relation. % In this model, moisture transport coefficients are assumed as the function of $ S $, such as permeability and resistance factor.

\subsection{Transport coefficients}
The resistance factor in Eq.~(\ref{eq:vapor_diffusion}) represents the reduction of accessibility for water vapor diffusion which is due to the presence of the solid and liquid phases, the tortuous path for diffusion, the different connectivities in the pore network, etc. Because of limited experimental results, the expression of $ f (S, \phi) $ is generally derived from theoretical concepts. For example, Millington and Quirk~\cite{Millington1961} deduced an equation for granular materials (soils):

\begin{equation} \label{eq:resistance_factor}
f (S, \phi) = \phi^{x_D} \left( 1 - S \right)^{x_D + 2}
\end{equation}

Millington and Quirk~\cite{Millington1961} proposed that parameter $ x_D $ was fixed at $ 4/3 $. However, granular materials are more porous than cementitious materials, so resistance to water-vapor diffusion may be more significant for cementitious materials. Thi\'{e}ry et al.~\cite{Thiery2008} suggested $ x_D $ = 2.74 based on the fitting of experimental data for cement pastes and mortars taken from Papadakis et al.~\cite{Papadakis1991}. %The comparison of $ f (S, \phi) $ calculated by these two proposed two values of $ x_D $ are shown in Fig.~\ref{fig:Resistance_factor_COCNCP} for the studied three materials. 
The comparison of $ f (S, \phi) $ calculated by these two proposed values of $ x_D $ shows that Thi\'{e}ry's suggestion provides smaller $ f (S, \phi) $ values (higher resistance) which may be closer to the real conditions of cementitious materials than the original $ x_D $ value for granular materials~\cite{Zhang2016}. % This suggestion seems more relevant for the range of porosity of cement pastes and mortars. 

%\begin{figure}[!ht]
%	\centering
%		\includegraphics[scale=0.3]{Resistance_factor_COCNCP.eps}
%\caption{Predicted resistance factor $ f (S, \phi) $ for the three studied cement pastes calculated by Millington's~\cite{Millington1959} and Thi\'{e}ry's~\cite{Thiery2008} formulas. }
%\label{fig:Resistance_factor_COCNCP}
%\end{figure}
%Validation of Van Genuchten–Mualem law for porous media has been verified for ordinary concrete~\cite{Kameche2014}. 

Another important transport coefficient is the relative permeability $ k_{rl} $. %The critical question about $ k_{rl} $ is whether hysteresis exists in $ k_{rl}(S) $. The independent domain theory provided by Poulovassilis~\cite{Poulovassilis1969} implies that $ k_{rl} $ in wetting should be larger than $ k_{rl} $ in drying for the same water content. This theory was supported by Mualem~\cite{Mualem1976b} who proposed a complicated hysteretic model for the prediction of $ k_{rl} $. However, experimental data for glass-beam (a very porous material) has shown there is no hysteresis in $ k_{rl}(S)$~\cite{Topp1966}. In these later studies this conclusion was further proved for sands and soils~\cite{Topp1971,Poulovassilis1971}. 
For cementitious materials, measuring the permeability to liquid-water for different RH is very difficult due to the fact that advective liquid transport and vapor diffusion always occur together; therefore, measured results include both transport mechanisms~\cite{VBB2007b}. Owing to these reasons, it is acceptable to assume that $ k_{rl} $ is a unique function of $ S $. One well-known model is the van Genuchten – Mualem equation (VGM) which was first reported by van Genuchten~\cite{VG1980}. It is formulated as a simple analytical relation 

\begin{equation} \label{eq:VGM}
k_{rl} (S) = S^{\ell} \left[ 1- \left(  1 - S^{1/m} \right)^{m} \right]^{2}
\end{equation}

In Eq.~(\ref{eq:VGM}), $ m $ is the same as in Eq.~(\ref{eq:VG2}). The term $ S^{\ell} $ is a correction factor which accounts for the influence of tortuosity. Different suggestions of parameter $\ell$ have been proposed by researchers~\cite{Mualem1976a, PL1987a}. In Mualem's research, $\ell$ varies between -1 and 3, and the value 0.5 was considered as the best choice. This value has also been used for cementitious materials~\cite{Mainguy2001,Zamani2014}.

%%%%%%%%%%%%%%%%%%%%
% NEW SECTION
%%%%%%%%%%%%%%%%%%%%
\section{Determination of liquid permeability}
\label{section:permeability} 
Methods chosen here for the determination of $K_l$ are primarily based on whether these methods can share the same specimen and whether this specimen can be used to obtain input and calibration data for the moisture transport model. The pivotal factor is the geometry of the specimen for these experimental methods. The KC and KTII equations need the crushed specimens to measure PSD, so they do not require a geometry for preparing specimens, and the same is true of the desorption isotherm measurements which only need small pieces. The ideal geometry for the drying experiments and the sorptivity measurements is a cylinder which is also convenient for performing 1D simulations. Owing to the large specimen size in the traditional flow-through methods, they suffer from the fact that the specimen can not easily be fully saturated.   
Considering the availability of equipment as well, we finally selected the following methods to determine $K_l$: beam bending, sorptivity, KC and KTII equations. 

\subsection{Beam bending}
Three-point beam bending has been developed as a method to measure the liquid permeability of a porous body~\cite{Scherer1992,Scherer2000a,Vichit2002,Vichit2003}. This method can provide permeability results within a few minutes to a few hours, whereas conventional techniques often require days and weeks. When a saturated porous material is bent, pressure gradients are created in the liquid, which flows within the pores to equilibrate the pressure. This phenomenon can be exploited to measure permeability, because a poroelastic analysis indicates the expected rate of change of the force $W(t)$ needed to sustain a constant deflection of the beam as the pore pressure relaxes \cite{Scherer1992}; by comparing the measured relaxation kinetics to the theoretical curve, the permeability $K_l$ can be extracted. 

% a rod of the saturated porous material is subjected to three-point bending. Because of the instantaneously applied strains, a pressure gradient develops in the liquid at the instant that the beam is deflected. As the liquid flows to restore ambient pressure, the force (W(t)) required to sustain a fixed deflection () decreases with time (t). The permeability (D) and Young’s modulus (Ep) can be extracted by fitting the measured relaxation data with the theoretical function.
When a saturated porous rod is subjected to bending, the flow in the porous medium is assumed to obey Darcy's law. The hydrodynamic relaxation is the process by which the liquid flow reestablishes ambient pressure throughout the specimen. The relaxation function $R(t)$ is given by normalizing the force exerted on the specimen $W(t)$ by the initial force $W(0)$. 

\begin{equation}\label{eq:BB_relaxation_force}
R(t)=\dfrac{W(t)}{W(0)}=1-A + A S_r(t)
\end{equation}
where the constant $A$ is

\begin{equation}\label{eq:BB_A}
A = \dfrac{\left(  \dfrac{1-2\nu_p}{3} \right) \left( 1- \dfrac{B_p}{B_s} \right)^2}{1- \dfrac{B_p}{B_s} + (1-\phi)\left( \dfrac{B_p}{B_l}- \dfrac{B_p}{B_s}\right) }
\end{equation} 
where $B=E(3(1-2\nu))$ is the bulk modulus. $\nu$ is the Poisson's ratio which is taken as 0.2 for cementitious materials~\cite{Vichit2002,Vichit2003}. Subscripts $p$, $l$ and $s$ represent properties of porous body, liquid phase and solid phase, respectively. For a cylindrical specimen, the relaxation function can be approximated as 

\begin{equation}\label{eq:BB_S}
S_r(t) = \exp \left[-\dfrac{8}{\pi^{1/2}} \left( \dfrac{\theta^{1/2}-\theta^{5/2}}{1-\theta^{1/2}} \right) \right]
\end{equation} 
where the reduced time is defined as 

\begin{equation}
\theta = \dfrac{t}{\tau_R}
\end{equation} 
where the hydrodynamic relaxation time $\tau_R$ is defined as (with the approximation $B_p/B_s \approx \phi^2$ for cement pastes~\cite{Vichit2002})

\begin{equation}\label{eq:BB_relaxation_time}
\tau_R = \left[\dfrac{2(1+\nu_p)}{3B_p} + \dfrac{1-\phi}{B_l}- \dfrac{1}{B_s}\left(\dfrac{\phi^2-5\phi+8}{5}\right)\right]  \left( \dfrac{\eta r^2}{K_l} \right)
\end{equation}
where $r$ (m) is the radius of the cylindrical rod. 

In the experiments, a constant displacement $\delta$ is suddenly applied to the rod within the linear elastic range and the force decay over time is continuously measured. The measured curve is then fitted by Eq.~\eqref{eq:BB_relaxation_force} with $A$ and $\tau_R$ as free parameters. From the plateau of the force relaxation curve, Young's modulus of the porous body can be calculated.

\begin{equation}
E_p =  \dfrac{L^3(1-A)W(0)}{12 \pi r^4 \delta}
\end{equation}
where $L$ (m) is the support span. 

According to the analytical solution for viscoelastic materials in~\cite{Scherer1992}, the total relaxation of the slender specimen is the product of the hydrodynamic $R(t)$ and the viscoelastic $\Psi_V (t)$ relaxation functions. 

\begin{equation}\label{eq:BB_total_relaxation}
\dfrac{W(t)}{W(0)} = R(t) \Psi_V (t) 
\end{equation}

This equation is valid for cases that the viscoelastic relaxation time $\tau_{V}$ is an order of magnitude longer than hydrodynamic relaxation time $\tau_R$. For a short-term measurement (e.g., a few hours), the viscoelatic relaxation function can be formulated by an exponential function. 

\begin{equation}
\Psi_V (t)  = \exp \left[-\left( \dfrac{t}{\tau_V} \right)^{b_V}  \right]
\end{equation}
where $b_V \subseteq \left[0, 1  \right]$ is a constant. The properties of liquid are taken from paper~\cite{Vichit2002}; thus, the fitting parameters can be determined and then used to extract $K_l$. 

\subsection{Sorptivity method}
Sorptivity is defined as \enquote{a measure of the capacity of the medium to absorb liquid by capillarity}~\cite{Philip1957}. 
It can be easily measured by simple experiments. Considering that the initial stage of mass changes in an absorption test is mostly controlled by capillary suction, sorptivity $S_p$ (m/s$^{1/2}$) can be determined by the cumulative volume of water crossing the specimen surface~\cite{Hall2011}

\begin{equation}\label{eq:sorptivity_measure}
\dfrac{\Delta m}{A_r\rho_l} = S_p t^{1/2} + a
\end{equation} 
where $\Delta m$ (g) is the measured mass change of the specimen, $A_r$ (m$^2$) is the area of the specimen's cross-section, and $a$ is a parameter associated with the end effect (such as buoyancy, lateral invasion). 

Using sorptivity directly to determine permeability is rarely discussed in the literature. Nevertheless, the determination of water diffusivity from sorptivity has been studied for several decades~\cite{Philip1960,Brutsaert1976,Parlange1994}. The relation between water diffusivity $D_l$ and permeability $K_l$ is 
 
\begin{equation}\label{eq:intrinsic1}
K_l = D_{l0} \mu \theta_s \left|\dfrac{\mathrm{d}S_l}{\mathrm{d}P_l}\right|_{S_l=1}
\end{equation}
where $D_{l0}$ (m$^2$/s) and $ \theta_s\left|\frac{\mathrm{d}S_l}{\mathrm{d}P_l}\right|_{S_l=1}$ are water diffusivity and water capacity of the porous material at saturated condition ($S_l=1$), respectively, and $g$=9.81 m/s$^2$ is the gravitational acceleration. 

For cementitious materials, Zhou~\cite{Zhou2014general,Zhou2016indirect} proposed the following equation for $D_{l0}$ 

\begin{equation}\label{eq:Zhou_solution_Dl0}
D_{l0} = {\tau_D}{\exp(n)} \left(\dfrac{S_p(\theta_i)}{ \theta_s - \theta_i  }\right)^2 
\end{equation}
where $\theta_s$ and $\theta_i$ are saturated and initial water contents, respectively, and $n$ is a shape parameter related to the initial saturation $S_{l,0}$ and can be expressed as $n=n_0(1-S_{l,0})$ ($n_0$ is the shape parameter for initially dried specimen). The coefficient $\tau_D$ can be either fitted by experimental data or calculated by~\cite{Lockington1999}

\begin{equation}\label{eq:Zhou_solution_tau}
\tau_D =\dfrac{n^2}{\left( 2n-1 \right) \exp(n) - n+1}
\end{equation}
%where $n$ is a shape parameter associated with the initial condition. 

The water capacity $\theta_s\frac{\mathrm{d}S_l}{\mathrm{d}P_l}$ is generally calculated from the sorption isotherm, while most isotherm equations show that the water capacity at saturated condition is infinity (e.g., Eq.~\eqref{eq:VG2}) or zero, which leads to meaningless $K_l$ in Eq.~\eqref{eq:intrinsic1}. Zhou~\cite{Zhou2014predicting} proposed a new equation for sorption isotherms.

\begin{equation}\label{eq:Zhou_S_l}
S_l = \left[ 1- c_1 + c_1 \exp\left( \dfrac{P_c}{c_2} \right)  \right]^{-1}
\end{equation}
where $c_1$ and $c_2$ (Pa) are two fitting parameters. Thus, $\frac{\mathrm{d}S_l}{\mathrm{d}P_l}$ is the derivative of this function at $S_l=1$.

\begin{equation}\label{eq:Zhou_C}
\left|\dfrac{\mathrm{d}S_l}{\mathrm{d}P_l}\right|_{S_l=1} = \dfrac{c_1}{c_2}
\end{equation}

Putting Eqs.~\eqref{eq:Zhou_solution_Dl0} and~\eqref{eq:Zhou_C} back to Eq.~\eqref{eq:intrinsic1}, $K_l$ can be calculated by the measured sorptivity $S_p$.

\subsection{Katz–Thompson (KT) equations}
Many attempts have been made to link the transport properties of porous media and their microstructure. One of the theories that have been widely used is Katz-Thompson (KT) theory which was initially developed to predict the permeability of sedimentary rocks. The percolation theory was employed in KT relation, that introduces the characteristic length as one of main inputs (KTI)~\cite{KT1986}. 

\begin{equation}\label{eq:KTI}
K_{l} = \dfrac{d_c^2}{226} \left( \dfrac{\sigma}{\sigma_0}\right)
\end{equation}
where $d_c$ is the characteristic dimension of pore space, which corresponds to the peak in the derivative of PSD, $\sigma$ is the electrical conductivity of the saturated porous material and $\sigma_0$ is the conductivity of pore solution. The coefficient 1/226 is used for general porous mateirals, but for concrete different values were suggested (e.g., 1/8 for the lightweight concrete~\cite{Sanchez2014study}). In this study, the modified values are not used because 1/226 is an analytical constant by assuming cylindrical pores in the MIP measurements~\cite{Ma2014mercury}, so it should not vary with materials. The conductivity was used to reflect the tortuosity of the pore network but it must be measured separately. Katz and Thompson~\cite{KT1987Prediction} proposed an expression for the conductivity term ($\sigma/\sigma_0$) which can be estimated from MIP data (KTII). 

\begin{equation}\label{eq:KTII}
\dfrac{\sigma}{\sigma_0} = \dfrac{d_{max}^e}{d_c} \phi V(d_{max}^e)
\end{equation}
where $d_{max}^e$ is the electrical conductivity characteristic dimension that produces the maximum conductance. For a very broad PSD, $d_{max}^e$ is estimated by $d_{max}^e = 0.34d_c$~\cite{KT1987Prediction}. $V(d_{max}^e)$ is the fractional volume of connected pore space with pore size larger than $d_{max}^e$. $d_c$ is normally taken as the critical (or breakthrough) pore diameter, representing the minimum radius which is geometrically continuous throughout hydrated cement paste. 

By using Eq.~(\ref{eq:KTII}), without adjusting parameters, Katz and Thompson concluded that permeability and $\sigma/\sigma_0$ can be predicted from the same MIP data~\cite{KT1987Prediction}. They found that the calculated $\sigma/\sigma_0$ showed good agreement with the directly measured values for sedimentary rocks.

\subsection{Kozeny–Carman (KC) equation}
%A good model to predict the permeability normally includes at least three parameters to account for the relationships between the flow rate and the porous space, such as the size of the pores, their tortuosity, and their connectivity. A frequently quoted relationship was proposed by Kozeny (1927) and later modified by Carman (1937, 1956). The resulting equation is largely known as the Kozeny–Carman (KC) equation. This equation was developed after considering a porous material as an assembly of capillary tubes for which the Navier–Stokes equation can be applied.  Since its first appearance to the present, this equation has been modified into several forms. One is commonly used:
%The KC equation was derived based on hydraulic radius theory for straight circular tubes. 
Considering that the flow in one-size straight tubes obeys Navier–Stokes (N-S) equation, Hagen–Poiseuille equation is an exact solution to the N-S equation. Meanwhile, Darcy's law can give the flux through these tubes. The comparison of Hagen–Poiseuille equation with Darcy's law yields water permeability $K_l$ as a function of the porosity $\phi$, the specific surface $S$ (m$^2$/kg), and tortuosity of channels $\tau$. One widely used version was proposed by Walsh and Brace~\cite{Walsh1984}. 

\begin{equation}~\label{eq:KC_1}
K_{l} = \dfrac{\phi^3}{2\tau^2 S^2 \rho_s^2}
\end{equation}
where $\rho_s$ (kg/m$^3$) is the bulk density of the dried material. 
%According to many classical soil mechanics textbooks (e.g., Taylor 1948; Lambe and Whitman 1969), the KC equation is approximately valid for sands and is not valid for clays. In practice, it was also claimed that the specific surface for soils is difficult to measure. Actually, Kozeny initially proposed the model to calculate the specific surface. 
%
%Another version of KC equation can be also found in the literature. 
%\begin{equation}
%k_{KC} = \dfrac{\phi d_h^2}{32 \tau}
%\end{equation}
%where $d_h$ is defined as the hydraulic diameter, which can be calculated based on MIP data $d_h = \frac{4V}{S}$ ($V$ is the total intrusion volume and $S$ is total pore area). 
As stated above, Eq.~\eqref{eq:KC_1} is developed for tubes having the same cross-sections. To account for the various sizes of pores, an equation adapted by Walsh and Brace~\cite{Walsh1984} can be used here. 

\begin{equation}~\label{eq:KC_2}
\dfrac{\phi}{\tau^2} =  \dfrac{\sigma}{\sigma_0}
\end{equation}
where the term $\sigma/\sigma_0$ is formulated in Eq.~\eqref{eq:KTII}. A similar equation for cementitious materials was also proposed by Wong et al.~\cite{Wong2006} who introduced a constrictivity factor and proposed a modified equation for permeability, but they found that the equation largely overestimated permeability. 

By plugging Eq.~\eqref{eq:KC_2} into Eq.~\eqref{eq:KC_1}, we thus have the version of KC equation that will be used in this study. 

\begin{equation}~\label{eq:KC_final}
K_{l} = \dfrac{\phi^2}{2 S^2 \rho_s^2} \left( \dfrac{\sigma}{\sigma_0}\right)
\end{equation} 

Since the tortuosity and porosity factors are included in this equation, it is supposedly representative of the pore structure in a porous material.

%%%%%%%%%%%%%%%%%%%%
% NEW SECTION
%%%%%%%%%%%%%%%%%%%%
\section{Experiments}
\label{section:experiment}
Measurements can be affected by many factors, such the type of cement, mixture procedure, the size of specimens, curing condition, etc. To avoid these artificial factors, we prepared all specimens in one batch. For most measurements (beam bending, sorption isotherm, drying, MIP, sorptivity), specimens are obtained from the same cylinder to minimize the influence of batch differences. 
The cement used in this study was a Type I ordinary Portland cement (OPC) from Buzzi Unicem, USA. Chemical and mineralogical data were given in~\cite{Zhang2017SEM}. 

The specimen preparation basically followed the procedure reported in~\cite{Vichit2002}. The cement paste with water-to-ratio of 0.5 is used in this study. After adding deionized water into cement powder, the material was hand mixed for 1 minute and then mixed by a vortex mixer (Stuart Vortex Mixers SA8) for additional 3 min. Before casting paste in the polystyrene pipettes, which were lubricated by petroleum jelly, the fresh paste was deaired for about 4 min. Two sizes of cylindrical specimens were prepared, 8 and 16 mm in diameter. After 48 h curing, cylinders were removed from the pipettes, wiped with lab tissues, and stored in limewater for further curing. 

\subsection{Measurements for $K_l$ methods} 
\subsubsection{Beam bending}
After one year, both 8 and 16 mm-diameter specimens were subjected to the beam bending measurements. Specimens were taken out and pressurized in a limewater-filled tube for one day to ensure that specimens were fully saturated prior to the tests. The test procedure has been well illustrated in the literature~\cite{Vichit2002,Vichit2003,ZhangJ2012} and a brief introduction is given here. The pushrod, which directly touched the slender specimen during tests, was controlled by a stepper motor to apply the sudden displacement. A linear-variable differential transformer (LVDT) was used to measure the deflection. A 250 g load cell was used to measure the load force. Since the relaxation is most rapid in the beginning, the data were recorded at logarithmic time intervals. The slender specimen was placed in a stainless steel container which was filled with limewater. Two V-shape steel supports were used to ensure good alignment of the cylindrical specimen and to avoid any movement during tests. The assembly was placed in an incubator to maintain constant temperature. 
The displacement applied depends on the radii of the specimen $r$ and the support span $L$. To have accurate results, the span $L$ must be longer than ten times the specimen diameter ($L>20r$). 

\subsubsection{Sorptivity}
\begin{figure}[!ht] 
	\centering 
    \includegraphics[width=7cm]{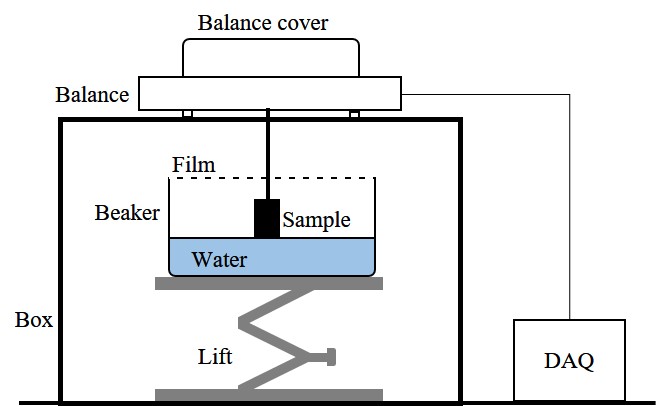}
\caption{Illustration for sorptivity measurements. }
\label{fig:sorptivity_design} 
\end{figure}

After BB tests, one of the 16 mm-diameter cylindrical specimens without any defects (mainly air voids, which can only be checked after cutting) was used to prepare specimens for other measurements. About 5 mm from each end of the cylinder was removed and discarded. The remaining part was gently cut by a diamond saw into many slices (about 1 mm thick) and several 20 mm long cylinders. Some short cylinders were used for the measurement of sorptivity. These cylinders were preconditioned for three months in cups with constant RHs (97\%, 85\%, 63\% and 53\%) which were achieved by using saturated salt solutions (see Table~\ref{table:salt_solution} for details). In the sorptivity measurements, only one end of the short cylinder can contact water. To minimize moisture exchange between the specimen and its surroundings, the side of the cylinder was sealed with the adhesive aluminum sheet and the other end was loosely covered to let gas escape when water penetrates from the opposite end. About 1 mm of the side surface at the end that contacts water was not sealed by the aluminum sheet. This can avoid the aluminum sheet contacting water and creating errors during measurements. 
An electronic balance (Denver Instrument) with accuracy $\pm$ 0.001 g was connected with a data acquisition system (DAQ) to automatically record the mass change (see Fig.~\ref{fig:sorptivity_design} for details). The top of the balance was covered to eliminate the air flow effect in the lab. A screwed fixture under the balance was used to grasp the specimen. Data recording every second started just after the specimen was fixed. A beaker with water was seated on a small lift that can be manually elevated until water just touches the bottom end of the specimen. The beaker was large enough (about 30 cm in diameter) to ensure that the drop of water level during measurements can be neglected. The beaker was covered by the plastic film and only a hole in the center was left for the specimen to pass through. Because the ambient RH is different to RHs for preconditioning, the time between taking the specimen out the cup and the beginning of the test should be as short as possible. Data logging continuously ran about 20 h but only the measured mass change curve at the initial stage was used to calculate sorptivity. 

%This study did not follow the standard sorptivity test since the specimen size is much smaller. 
%specimen conditioning flows the standard ASTM 1585. First, the specimens were placed in the sealed chamber with RH of 80\% at 50$^\circ$C for one week and then transferred to another container at room temperature. At 50$^\circ$C, the vapor diffusion coefficient is about 4 times higher than that at 25$^\circ$C~\cite{Harris1980}, so this process can shorten the specimen conditioning. 

\subsubsection{MIP, NAD and TGA} 
A small part of the slender specimen was used to prepare specimens for MIP, NAD and TGA measurements. As reported in our previous studies, different drying methods have significant effects on the microstructure~\cite{Zhang2017SCD,Zhang2017SEM}, so two drying methods were compared in this study: oven drying at 60$^\circ$C while flushing with N$_2$ and isopropanol (IPA) replacement followed by drying under flowing N$_2$ at room temperature. The details of these two drying methods can be found in~\cite{Zhang2017SEM}. Part of the rod was crushed into lumps (about 3 mm) which were subjected to different drying methods and fine powders were used for the first TGA test (labelled with \enquote{No drying}). 
After drying, lumps were further crushed and sieved. Particles with size $<$ 0.6 mm were used for the TGA measurements, size between 0.6 and 1.2 mm were used for the NAD tests and size about 3 mm were used for the MIP measurements. 
TGA measurements were performed by PerkinElmer\textsuperscript{\textregistered} Pyris 1, in which specimens were heated from room temperature to 1000 $^\circ$C. The measured mass change between 105 $^\circ$C and 1000 $^\circ$C was used to calculate the degree of hydration (DoH). The detailed procedure and equation were reported in our previous studies~\cite{Zhang2011,Zhang2017SCD,Zhang2017SEM}. 
The NAD measurements were conducted using an ASAP 2010 apparatus, from Micromeritics (see~\cite{Zhang2017SCD} for more details). 
The Micromeritics 9410 apparatus was used for MIP tests (see~\cite{Sun2010pore} for details). 

%MIP can be only done on dried specimens. Two drying methods were used here: IPA replacement followed by flowing N2 drying and direct flowing N2 drying at 60C. The N2 flow can provide a zero-RH and CO2 free environment. 
%The specimen size is about 3 mm in diameter (check with Ma's paper). 
%

\subsection{Measurements for moisture transport model} 
\subsubsection{Sorption isotherm}
The desorption isotherm was measured by using saturated salt solutions to establish ten different RHs (see Table~\ref{table:salt_solution}). At each RH, three thin slices (initially saturated) with thickness around 1 mm, which were cut from the 16 mm-diameter cylindrical specimen, were used to determine the water content. There was about 50 ml saturated salt solution in a 200 ml cup and a plastic mesh was installed in the middle that specimens were placed on. There was a small hole in the lid of the cup, so that the hang wire could pass through and the lower end was hooked to a specimen pan containing one slice. A rubber stopper was used to seal the hole in the lid between measurements. When measuring the mass of the slice, one just needs to remove the rubber stopper and connect the upper end of the hang wire with the electronic balance. Thus, the mass change of the specimen could be regularly monitored with minimal disturbance caused by opening the cup. Note that water content at different RHs was measured by using different specimens, this can greatly reduce measuring time compared with the stepwise method for the same specimen (e.g.,~\cite{VBB2007a}). 

%Because of the complexity of sorption processes, the isotherms cannot be determined by calculation, but must be recorded experimentally for each product. 
%
%capillary saturation 
%The saturated state was reached by capillary suction, meaning that the isolated pores are not considered in the calculation of $S$, regardless of whether they are empty of filled with water. 
%
%there are significant differences between equilibrium (stepwise) drying and drying directly to a very low RH.

\begin{table}[!ht]
\small
\centering 
\caption{Salts used to control different RHs.} 
\begin{tabular}{ p{1cm} p{1.2cm} p{1.2cm}p{1.2cm}p{1.4cm}p{1.2cm}p{1.2cm}p{1.2cm}p{1.2cm}p{1.2cm}p{1.2cm}  }  
\hline  
RH \% & 11 & 22 & 33 & 44 & 53 & 63 & 75 & 85 & 92 & 97    \\ %\hhline{~-------}
Salt & Lithium Chloride & Potassium Acetate & Magnesium Chloride & Potassium Carbonate & Magnesium Nitrate & Sodium Bromide & Sodium Chloride & Potassium Chloride & Potassium Nitrate & Potassium Sulfate   \\ %\hhline{~-------}
Formula & LiCl & KCH$_3$COO & MgCl$_2$ & K$_2$CO$_3$ & Mg(NO$_3$)$_2$ & NaBr & NaCl & KCl &KNO$_3$ & K$_2$SO$_4$ \\ \hline 
\end{tabular}\label{table:salt_solution}
\end{table}

\subsubsection{Drying tests}
Drying tests at RH = 22, 53, and 63 \% were performed on the short cylindrical specimens that were first preconditioned at RH=97\% rather than starting from the saturated condition. The preconditioning was done by putting the initially saturated short cylinders in a cup with 97\% RH and keeping at 40 $^\circ$C for about three days. The cup was then moved to the room temperature for one month. The high temperature used here can speed up water evaporation and reduce preconditioning time. After preconditioning, it was assumed that moisture uniformly distributed in the specimen. The cylinders were then sealed by the adhesive aluminum sheet on the side and only two ends were exposed to the corresponding RH environment. The cups with the same design as the desorption isotherm measurements were used for drying tests. The masses of these specimens were regularly weighted for about three months. 

After tests, some short cylinders (from sorptivity measurements and drying tests) were dried in an oven at 60 $^\circ$C to obtain the mass at dryness. Since the geometry of these cylinders were well defined, the mass difference between the dried and saturated states can be used to determine the porosity. 

%%%%%%%%%%%%%%%%%%%%
% NEW SECTION
%%%%%%%%%%%%%%%%%%%%
\section{Results}
\label{section:results}

\subsection{$K_l$ determined by different methods}
\begin{figure}[!ht] 
	\centering 
		\includegraphics[width=7cm]{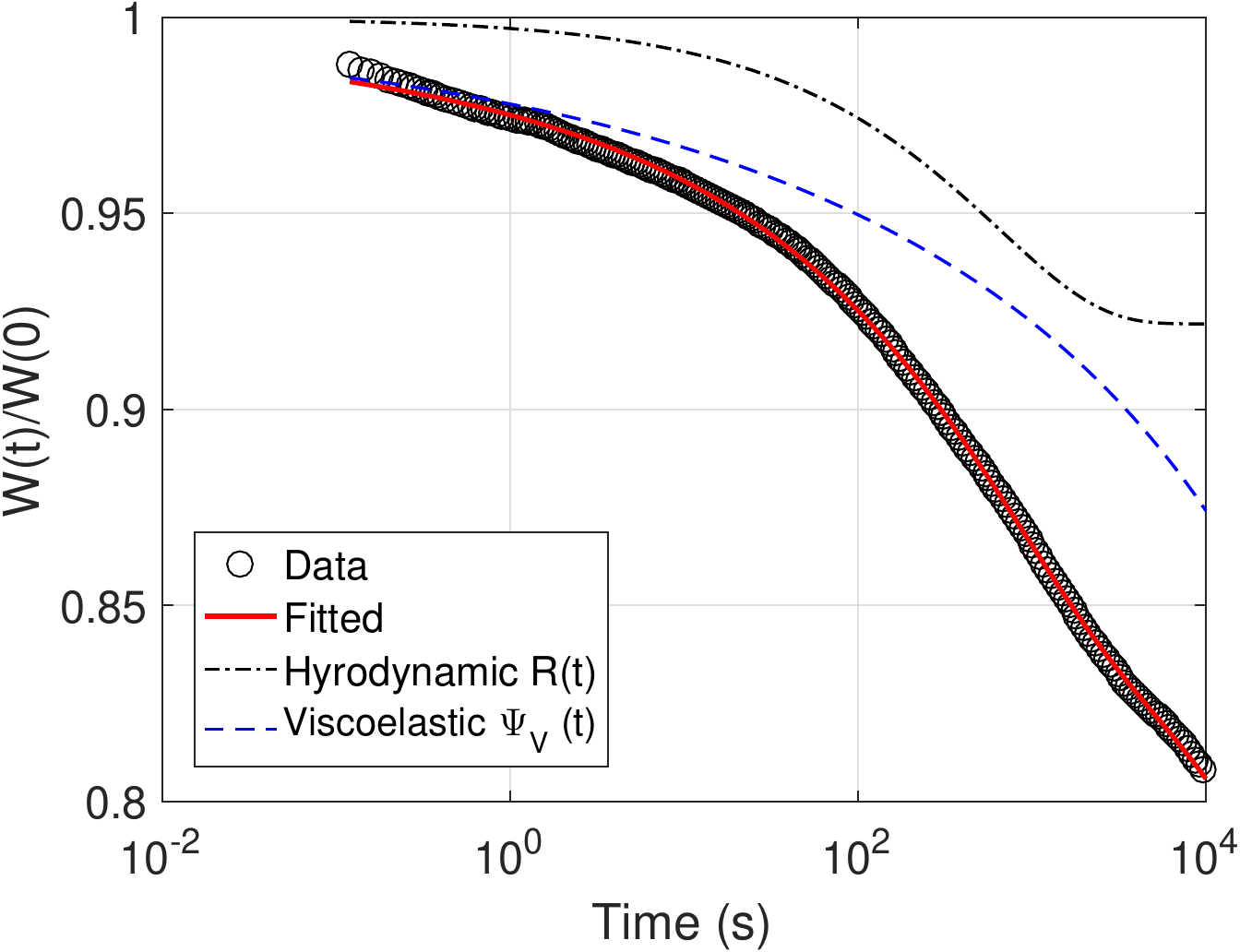}
\caption{Measured and fitted relaxation curves for BB tests ($2r$=16 mm). }
\label{fig:result_BB}
\end{figure}

% BB 
Specimens with two diameters (8 and 16 mm) were used in the beam bending tests and one example for the larger specimen is shown in Fig.~\ref{fig:result_BB}. Good agreement between measured and fitted relaxation curves can be seen in this figure with well separated hydrodynamic and viscoelastic relaxation curves. The inflection point represents the end of the hydrodynamic relaxation and is clearly shown at about 1000 s. This time is much delayed compared with those reported in~\cite{Vichit2002,Vichit2003} because the material in the present study is older. Consequently, the measured permeability (shown in Table~\ref{table:result_BB}) is lower than those in~\cite{Vichit2002,Vichit2003}.

Table~\ref{table:result_BB} shows data form the BB tests for two specimens. If the flow is radial, Eq.~\eqref{eq:BB_relaxation_time} indicates that the hydrodynamic relaxation time depends on the square of the radius. As shown in Table~\ref{table:result_BB}, $\tau_r$ for the 16-mm specimen is about 4 times the average $\tau_r$ for 8-mm specimen\footnotemark[1]. This means that the results scaled with the specimen size are in agreement with the theory. 

\begin{table}[!ht]
\small
\centering 
\caption{Parameters in the beam bending method for different size of specimens.} 
\begin{tabular}{ p{2.5cm}p{1.2cm}p{2.5cm}p{2.5cm}}  \hline  
Specimen diameter & $E_p$(GPa)\footnotemark[2] & $\tau_R$(s) & $K_l$ ($10^{-21}$m$^2$)   \\ \hhline{----}
16 mm   & 23  & 20410 & 0.679   \\ \hhline{----}
8 mm(1st) & 23 & 8469 & 0.4   \\ \hhline{----}
8 mm(2nd) & 23 & 2189 & 1.4   \\ \hhline{----}
\end{tabular}\label{table:result_BB}
\end{table}
\footnotetext[1] {It generally plots the $\tau_R$ vs. $r^2$ curve, which shows that two tests for 8 mm specimens scatter around a line that passes through the origin, but the average value of $\tau_R$ is on the same line with the 16 mm specimen. } 
\footnotetext[2] {The Young's modulus is taken from~\cite{Constantinides2004effect} which reported 22.8$\pm$0.5 GPa for the mature cement paste. In fact, the value of $E_p$ does not affect the calculated permeability which is mainly dependent on the inflection point of the measured force relaxation curve. $\tau_R$ is the parameter controlling the position of the inflection point. For cementitious materials, in which $B_p$ and $B_s >> B_l$~\cite{Scherer2007a}, Eq.~\eqref{eq:BB_relaxation_time} can be approximately replaced by $\tau_R \approx \left(\dfrac{1-\phi}{B_l}\right)  \left( \dfrac{\eta r^2}{K_l} \right) $. It is clear that $K_l$ is primarily dependent on the mechanical property of liquid.} 

\begin{figure}[!ht] 
	\centering 
		\includegraphics[width=7cm]{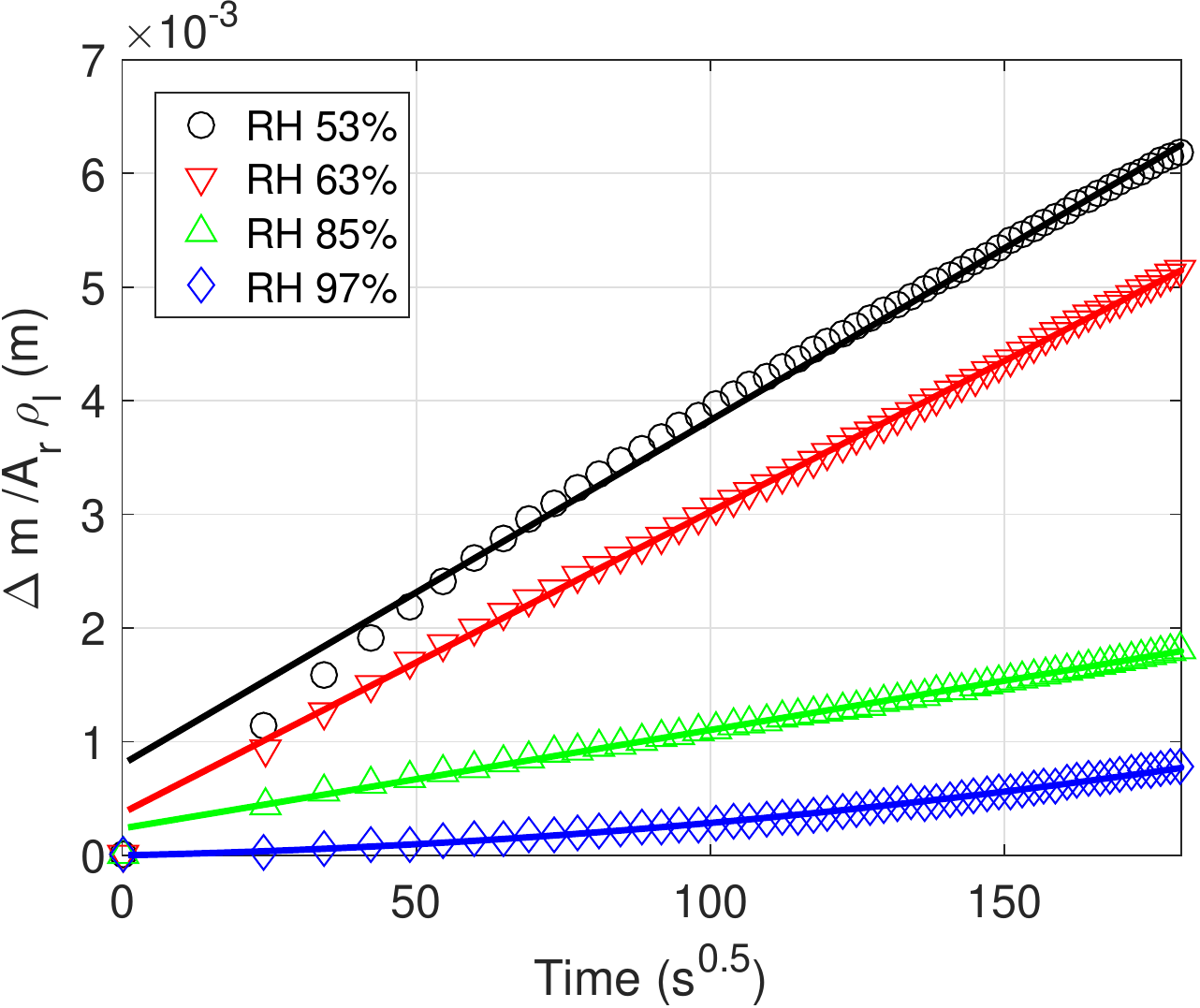}
\caption{Measured and fitted sorptivity experimental curves (symbols are measured and lines are fitted). The number of measured data is reduced to clarify. }
\label{fig:result_sorptivity}
\end{figure}

% sorptivity 
Even though the sorptivity tests ran about 20 h, after a certain period (about 10 h in this study), the rate of mass increase started to drop because the water penetration front reached the top of the specimen. At the initial stage, water uptake is supposed to be a nearly linear function of the square root of time as indicated in Eq.~\eqref{eq:sorptivity_measure}. In this study, we only took the linear part of the measured curves to fit sorptivities for specimens with various initial water contents. Fitted and measured results are compared in Fig.~\ref{fig:result_sorptivity} showing very good fittings. 

The fitted sorptivities for all tests are given in Table~\ref{table:result_sorptivity} which clearly shows the decease of $S_p$ with the increase of the initial water content. This is very reasonable when the specimen has less water in the pore network (large pores are not filled with liquid water), the initial capillarity is much stronger than specimens having more water inside, and thus water uptake is much faster at the initial stage. The parameter $n_0$ in Eq.~\eqref{eq:Zhou_solution_Dl0} is a material-dependent parameter. Since only one material is studied here, we take $n_0=6$ for all tests~\cite{Lockington1999}. Note that this value is smaller than those in~\cite{Zhou2016indirect} (6.2 $\sim$ 7.6) for concretes. The calculated permeabilities based on measured sorptivities in Table~\ref{table:result_sorptivity} show a very small variation with all values in the same order of magnitude. 

A clear tendency is seen for initial RH from 53 - 85 \% as the calculated $K_l$ decreases due to the decrease of sorptivity. Nevertheless, $K_l$ for initial RH = 97\% is much higher these for the other initial RHs. The primary reason is that 97 \% RH only can remove water in the large capillary pores ($>$ 35 nm according to Young-Laplace and Kelvin equations), so the small capillary and gel pores are not affected under this drying condition. Theoretically, water transport in all pores should be included in the measurements; this is, a completely dried specimen is more representative than a partially dried specimen, while drying can change the virgin microstructure. The compromise between drying condition and microstructural change must be well analyzed prior to performing sorptivity measurements. In addition, we found that $K_l$ for initial 97 \% RH is very sensitive to the choice of sorption isotherms. For instance, $K_l$ determined by using the measured saturation is about 2 times greater than that by using saturation calculated by the VG equation. Therefore, results presented in Table~\ref{table:result_sorptivity} use the measured sorption data. This also implies that the use of specimens with high initial water content is inappropriate for the sorptivity method. 

\begin{table}[!ht]
\small
\centering 
\caption{Results from sorptivity tests.} 
\begin{tabular}{ p{2.5cm} p{1.2cm} p{1.2cm}p{1.2cm}p{1.2cm}}  
\hline  
RH (\%) & 53 & 63 & 85 & 97   \\ \hhline{-----}
$S_p$ ($10^{-5}$m/s$^{0.5}$)  & 3.03 & 2.56  & 0.866 & 0.108   \\ \hhline{-----}
$K_l$ ($10^{-21}$m$^2$)\footnotemark[3] & 165 & 153 & 104 & 503   \\ \hhline{-----}
\end{tabular}\label{table:result_sorptivity}
\end{table}
\footnotetext[3] {Saturation and water content in Eq.~\eqref{eq:Zhou_solution_Dl0} are taken from the measured sorption isotherm rather than calculated by the VG or Zhou's equations. }

%\begin{figure}[!ht] 
%	\centering 
%		\includegraphics[scale=0.5]{figures/Kl_sorptivity_Zhou.eps}
%\caption{Permeability calculated by sorptivity data and fitted by a linear curve. }
%\label{fig:result_Kl_Zhou}
%\end{figure} 

% KC and KT 
Results from MIP, NAD and TGA are provided in Table~\ref{table:result_KCKT} for specimens dried by different methods. 
After one year hydration, about 11\% cement is still unhydrated, which is largely due to the fact that the cores of some cement grains have less accessibility to water because of the shell around them and the complex pore network in cementitious materials. 
The specimen dried by N$_2$ at 60 $^\circ$C shows a slight lower DoH than the undried specimen. This is because 60 $^\circ$C leads to the decomposition of some hydration products (mainly ettringite) and the loss of chemically bond water~\cite{Zhang2017SCD}. 
The isopropanol replacement shows higher DoH than others. This difference could be attributed to stronger interactions of IPA (probably strong physical adsorption) with hydration products (mainly calcium hydroxide), so that IPA is gradually released during heating in the TGA~\cite{Zhang2017SCD}. %Unfortunately, we were not able to identify the carbonate-like phases even through we tried several characterization techniques~\cite{Zhang2017SCD}. This is probably because the carbonate-like phases are so few to be detected. 
Drying effects are also clearly shown in the surface area measured either by N$_2$ BET or by MIP. They both show that IPA replacement provides higher surface area than nitrogen drying at 60 $^\circ$C. It is generally believed that IPA replacement preserves the fine microstructure, while the capillary force induced by water evaporation during the flowing nitrogen drying can cause the collapse of the fine microstructure. 
In addition, the N$_2$ BET surface area is always higher than MIP as reported in the literature~\cite{Thomas1999surface}. This results from the fact that mercury is not able to enter the fine pores that need high intrusion pressure which may damage the fine pore structure. This difference is also illustrated by the porosity measured by the mass difference and MIP in Table~\ref{table:result_KCKT}. 

%Taylor [10], the "typical range" of SN2 for mature OPC paste is given as 10-150 m2/g of dry paste, but even higher results have been obtained after solvent exchange

%mercury intrusion porosimetry , which do not measure surface in the finest pores
%C-S-H gels adapt their morphology to the physical surroundings. 
%MIP, which is not sensitive to the smallest pores and therefore measures a much lower surface area than other techniques.

The permeability values in Table~\ref{table:result_KCKT} are calculated based on MIP data. NAD data are not used here because the calculated incremental PSD monotonically increases with the pore size and therefore the critical pore diameter can not be identified. $K_l$ provided by the KC equation is about three times as low as that by KT equation for both drying methods. The version of KT equation used in this study (KTII, see Eq.~\eqref{eq:KTII}) is much simplified compared to the original one (KTI in~\cite{KT1986}). KTII mainly relies on the measured critical pore size $d_c$ and therefore the accuracy and robustness of KTII are also lower. The modified KC equation, by contrast, including the effects of surface area and density of solids, may be more representative for porous properties of a material. %The density of solid $\rho_s$ can account the difference between various porous materials, . 

\begin{table}[!ht]
\small
\centering 
\caption{Results from MIP, NAD and TGA, and calculated $K_l$ by KC and KT methods.} 
\begin{tabular}{ p{2.5cm} p{1.2cm} p{1.5cm}p{1.2cm}p{1.2cm}p{1.2cm}p{1.2cm}p{1.4cm}p{1.4cm}}  
\hline  
Specimen     &  DoH  & $S$(N$_2$ BET) (m$^2$/g) & $S$(MIP) (m$^2$/g) & $d_c$ ($\mu$m) & $\phi$ & $\rho_s$\footnotemark[6] (g/ml) & $K_l$(KC) (10$^{-21}$ m$^2$) & $K_l$(KT) (10$^{-21}$ m$^2$)   \\ \hhline{---------}
No Drying  & 0.886 & -  & - & - & 0.358\footnotemark[4] & - & - & -    \\ \hhline{---------}
N$_2$ 60$^\circ$C  & 0.842 & 108  & 57.4 & 0.05 & 0.250\footnotemark[5] & 1.569 & 81.2 & 233    \\ \hhline{---------}
IPA replacement  & 0.899 & 124  & 64.7 & 0.04 & 0.244\footnotemark[5] & 1.602 & 50.7 & 134    \\ \hhline{---------}
\end{tabular}\label{table:result_KCKT}
\end{table}
\footnotetext[4] {Measured by the mass difference of a specimen between saturated and dried (60$^\circ$C) states.} 
\footnotetext[5] {Measured by MIP.}  
\footnotetext[6] {Calculated based on MIP data.}  

%KC equation in this study provides exactly same results with 
%\begin{equation}
%K_l = \dfrac{\phi d_a^2}{32 \tau}
%\end{equation} 
%where $d_a^2$ is the average pore size determined by MIP data (4V/S). Note the tortuosity $\tau$ here has different formula to Eq.~\eqref{eq:KC_2}. 

It is clear that these $K_l$ values determined by various methods show a large scatter even though all tests were based on exactly the same material. The permeability $K_l$ determined by the BB method is much lower than other methods. One possible reason is that the specimen used in the BB method is fully saturated and specimens are dried to some extend in the other methods. As reported in the literature~\cite{Fonseca2010,Zhang2017SCD}, any drying can alter the morphology of C-S-H gel and the microstructure of material to a certain degree. From this point, we expect that the BB method is able to provide $K_l$ much closer to the \enquote{true} one than the other methods. 
The permeabilities from KT and sorptivity methods are very close and higher than KC. The reason for the difference between the sorptivity method and the others is unknown to us, and more work needs to be done for the sorptivity method since the microstructural changes during water absorption in cementitious materials are not considered by the sorptivity method. 
%Either measured or predicted values always show a large scatter. Hence, if all predicted/measured permeability values can hold within an order of magnitude, we would see they are close.  

\subsection{Comparison of $K_l$ from different methods}
\begin{figure}[!ht] 
	\centering 
		\includegraphics[width=7cm]{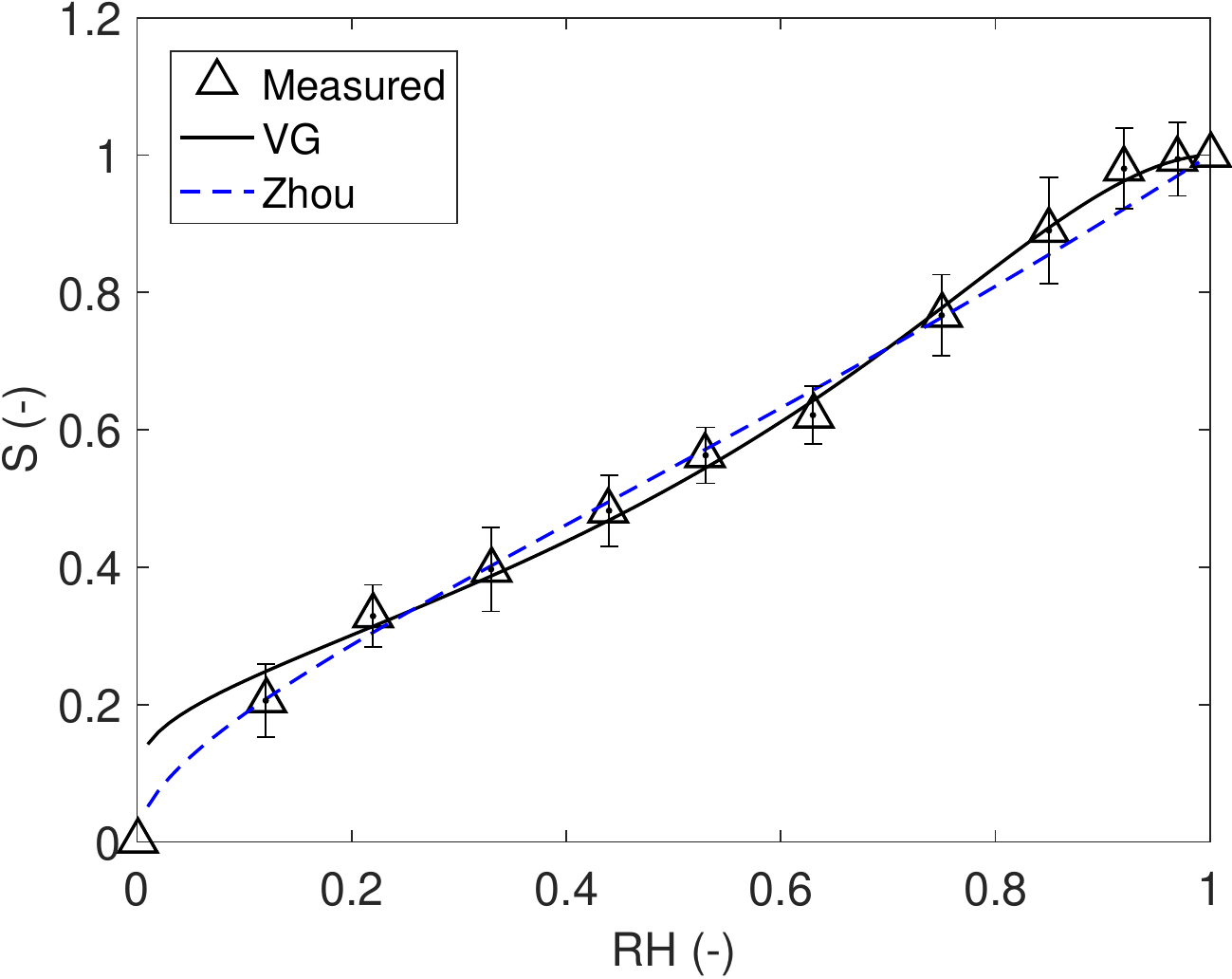}
\caption{Measured and fitted desorption isotherm. }
\label{fig:result_WVSIs}
\end{figure}

% sorption isotherm 
Measured and fitted desorption isotherms by Eqs.~\eqref{eq:VG2} and~\eqref{eq:Zhou_S_l} are shown in Fig.~\ref{fig:result_WVSIs}. Fitted parameters for Eqs.~\eqref{eq:VG2} and~\eqref{eq:Zhou_S_l} are provided in Table~\ref{table:WVSIs_parameters}. The calculated adjusted determination coefficient $R_{adj}^2$ (equation can be found in~\cite{Zhang2014,Zhang2017DT}) are very close to 1, indicating that both equations have a good applicability for the studied cementitious material. 

\begin{table}[!ht]
\small
\centering 
\caption{Fitted parameters for the measured desorption isotherm.} 
\begin{tabular}{ p{1.5cm} p{1.2cm} p{1.2cm}p{1.2cm}p{1.2cm}}  
\hline  
Equation  & \multicolumn{2}{c}{VG (Eq.~\eqref{eq:VG2})} & \multicolumn{2}{c}{Zhou(Eq.~\eqref{eq:Zhou_S_l})}    \\ \hhline{-----}
\multirow{2}{1.5cm}{Parameter} & $\alpha$ (Pa) & $m$ & $c_1$ (Pa) & $c_2$   \\ \hhline{~----}
 & 4.39E7 & 0.423 & 2.66E8 & 1.964   \\ \hhline{-----}
$R^2_{adj}$  & \multicolumn{2}{c}{0.994} & \multicolumn{2}{c|}{0.992} \\ \hline 
\end{tabular}\label{table:WVSIs_parameters}
\end{table}

In Fig.~\ref{fig:result_WVSIs}, the VG equation shows slightly better fitting at high RH, while Zhou's equation is slightly better at low RH. 

% comparison of measured and simulated mass loss curves 
In the moisture transport model introduced in Section~\ref{section:Modelling}, $K_l$ is the only unknown, so values calculated by the above methods can be used. By comparing the simulated mass loss curves using these $K_l$ values with the measured ones, we can tell which $K_l$ determination method is suitable for this moisture transport model. 
For the BB method, $K_l$ measured for the 16-mm specimen is used for simulation since the drying tests were done for the specimens with the same size. 
For KC and KTII equations, $K_l$ determined by the specimen dried by IPA replacement is chosen because we have shown that IPA replacement can better preserve the delicate microstructure of cementitious materials than the flowing nitrogen drying. 
For the sorptivity method, $K_l$ values determined from specimens preconditioned at RH = 53 and 63\% are directly used for drying tests with the corresponding RHs. For drying at RH = 22\%, there is no corresponding sorptivity measurement, so we take the $K_l$ value of the sorptivity measurement for initial RH = 53\% (1.65$\times 10^{-19}$ m$^2$). 

\begin{figure}[!ht] 
	\centering 
	\subfigure[Drying test at RH=22\%.]{
		\includegraphics[width=6cm]{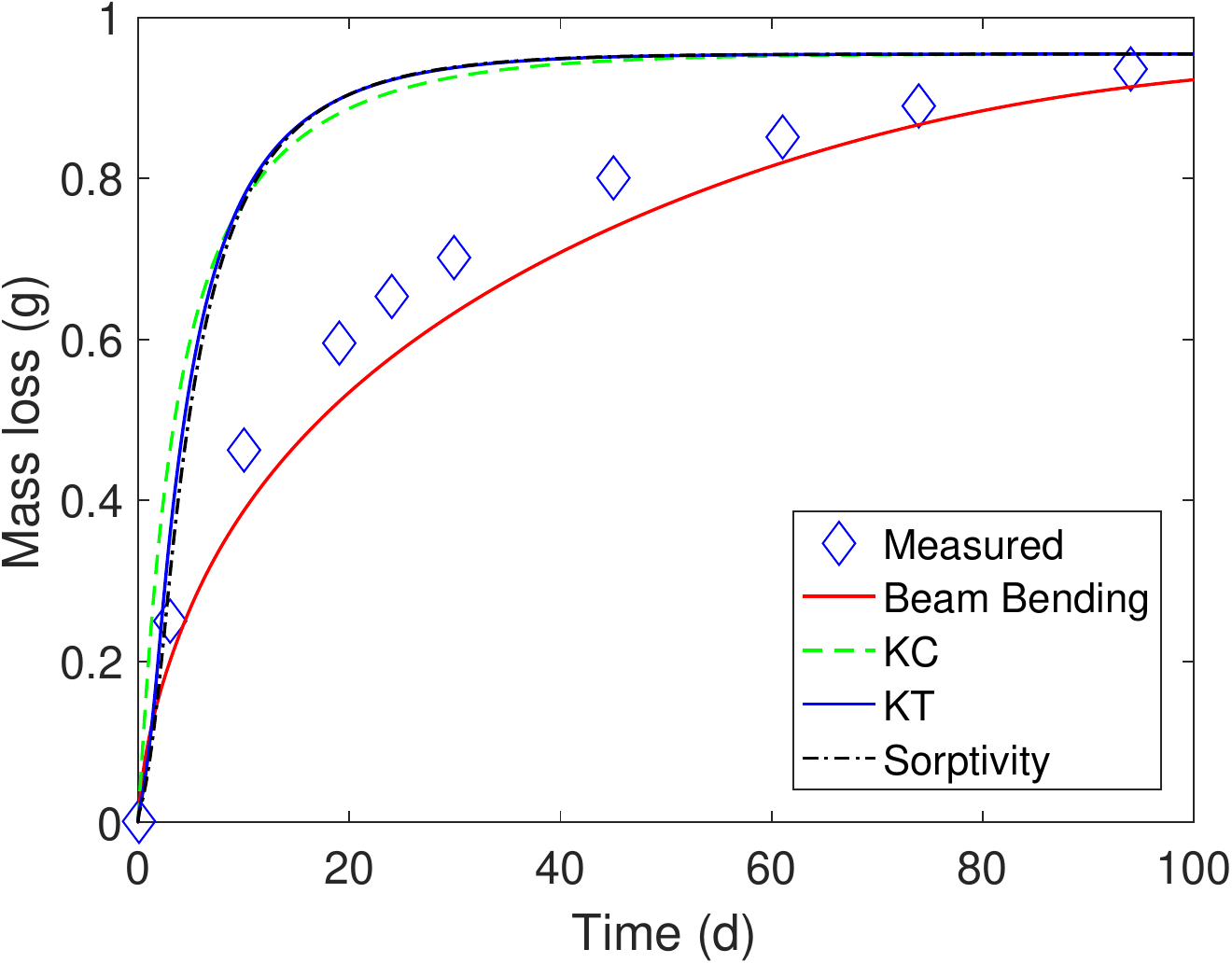}
		}\\
	\subfigure[Drying test at RH=53\%.]{
		\includegraphics[width=6cm]{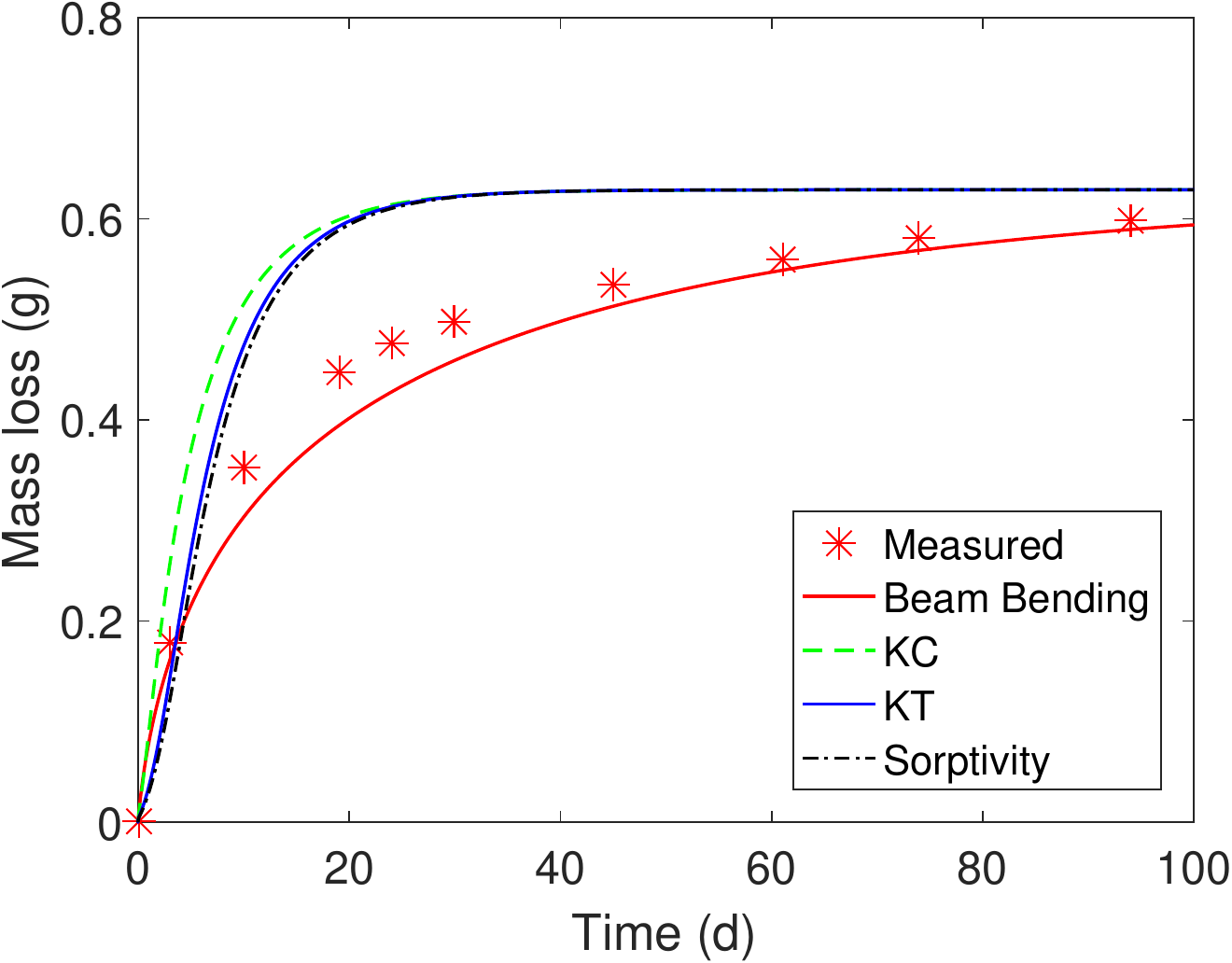}
		}\\
	\subfigure[Drying test at RH=63\%.]{
		\includegraphics[width=6cm]{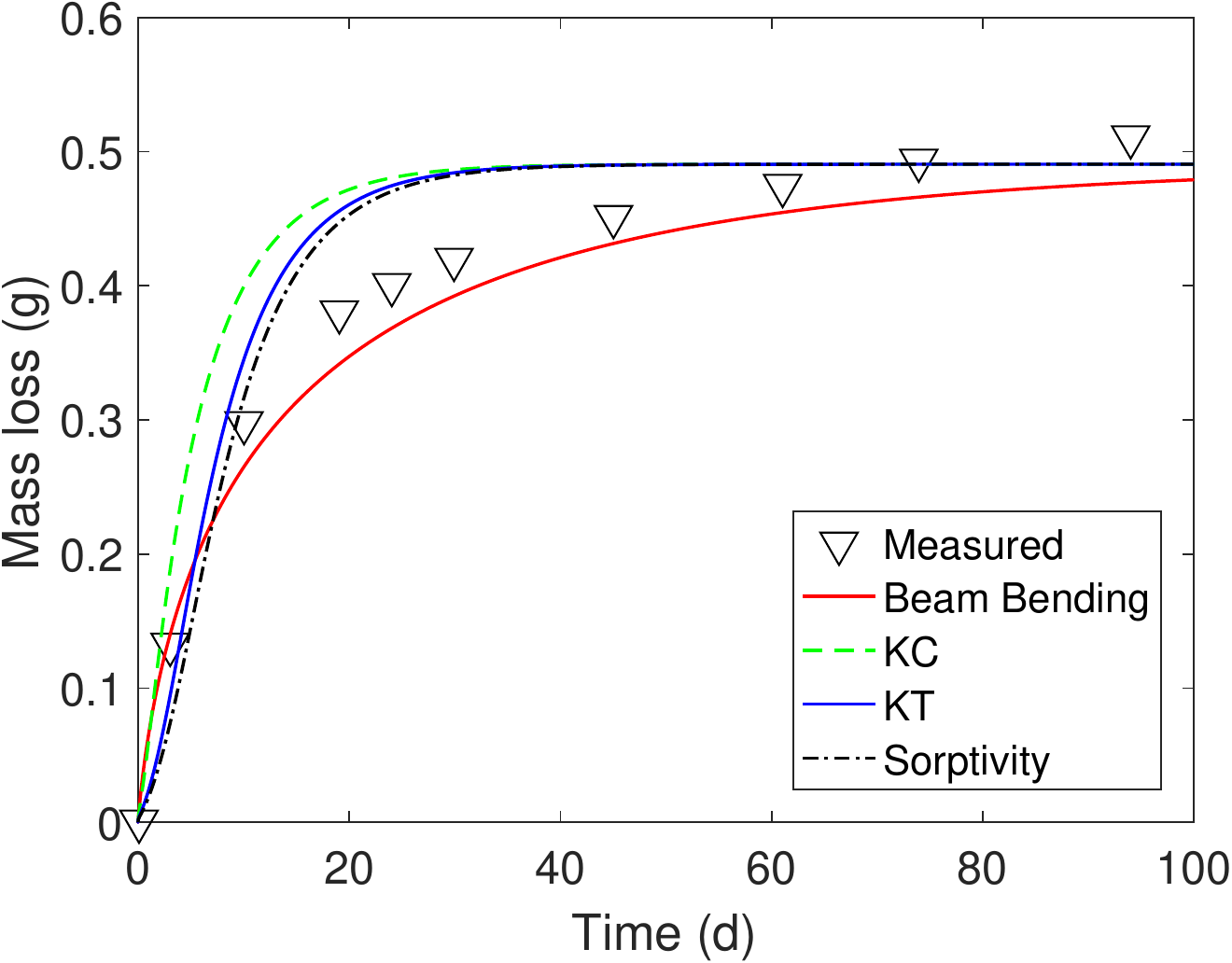}
		}
\caption{Comparison of measured and simulated mass loss curves.}
\label{fig:result_mass_loss} 
\end{figure}

Comparison of measured and simulated mass loss curves in Fig.~\ref{fig:result_mass_loss} clearly demonstrates that $K_l$ determined by the BB method is able to provide the curves closest to the measured ones regardless of the drying condition. Although permeabilities by KC, KTII and sorptivity methods are different, the mass loss curves calculated by them are very close, because these permeabilities are so high that they cause mass loss quickly reaching the plateau. Therefore, the similar conclusion as the literature can be drawn that KC and KTII equations overestimate $K_l$ by 1-2 orders of magnitudes \cite{VBB2011b,ElDieb1994,Zhou2017}

%%%%%%%%%%%%%%%%%%%%
% NEW SECTION
%%%%%%%%%%%%%%%%%%%%
\section{Discussion} 
%experimental data from~\cite{Kameche2014},~\cite{Monlouis-Bonnaire2004}

\subsection{Inverse analysis}
In fact, $ K_l $ in the moisture transport model can be determined by back calculation from the measured mass loss curve at a constant RH. This method was known as \enquote{inverse analysis} in the previous studies~\cite{Coussy2001, Mainguy2001,VBB2007c, Zhang2015,Zhang2016}. As discussed in~\cite{Zhang2015,Zhang2016}, the initial moisture state of calculations corresponds to the state after self-desiccation which is about the same RH used in this study for the preparation of specimens for drying tests. The previous study reported that the inversely determined $K_l$ for the same material dried at two RHs is very close~\cite{Zhang2015}, which proves that the moisture transport model used here is able to provide the consistent results for $K_l$. Moreover, water saturation profiles simulated by this moisture transport model match the measured ones very well. This further strengthens our confidence in using the moisture transport model in Section~\ref{section:Modelling} for inverse calculation of $K_l$.

\begin{figure}[!ht] 
	\centering 
		\includegraphics[width=7cm]{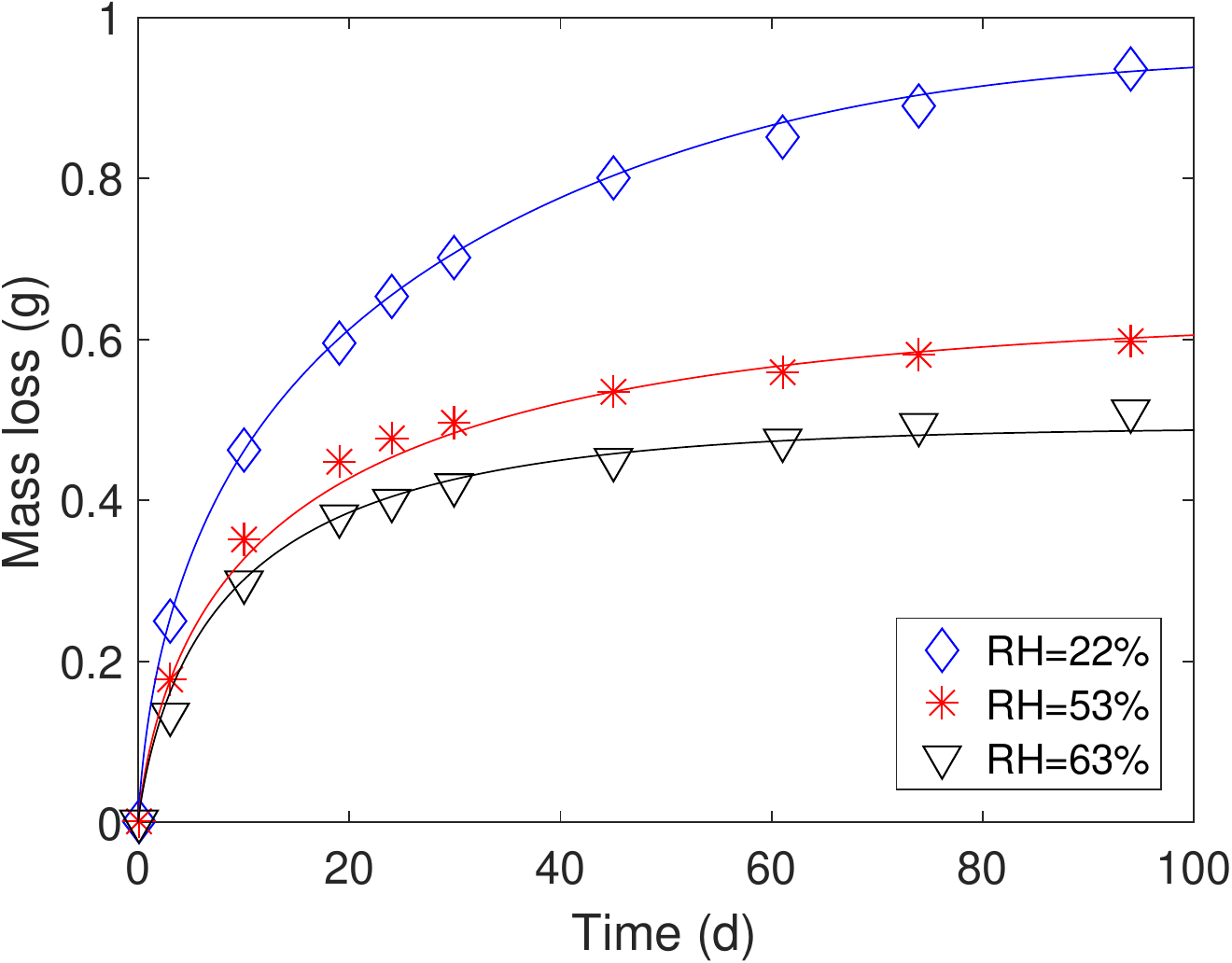}
\caption{Comparison of measured and simulated mass loss curves for drying at three different RHs. }
\label{fig:mass_loss}
\end{figure}

The fitted mass loss curves for drying at three different RHs are shown in Fig.~\ref{fig:mass_loss}. By only adjusting $K_l$ in the moisture transport model (of course boundary condition must be changed to the corresponding drying condition), we are able to obtain very good fitting with the measured curves. 
Permeabilities inversely determined by fitting the measured mass loss curves are given in Table~\ref{table:inverse_analysis}. 
The values of $ K_l $ for three different drying conditions are very close with difference less than a factor of two. 
Compared with permeabilities determined by the above methods, the inverse analysis shows $K_l$ values much lower than those from KC, KTII and sorptivity methods and slightly higher than that from the BB method. This confirms our hypothesis that the BB method, using the saturated specimen, is able to provide $K_l$ suitable for the moisture transport model employed here. 
As mentioned above, the good capability of the BB method is because the saturated specimen retains the original microstructure of cementitious materials while the microstructure of dried specimens in other methods is altered. There is no correction for the microstructural change in the KC and KTII equations. In the sorptivity method, the correction may be partially considered in the measured desorption isotherm. Nevertheless, during water absorption, the C-S-H gels undergo swelling, recovery and rearrangement~\cite{Jennings2008,Fischer2015observation,Zhou2017}, which are not taken into account in the sorptivity method. 
For the moisture transport model, the microstructural change is not directly considered by any equation, while the input data - the measured desorption isotherm - actually includes this information. Unlike the sorptivity method, the inverse analysis is not affected by the wetting of C-S-H gels.

\begin{table}[!ht]
\small
\centering 
\caption{Permeabilities determined by the inverse analysis.} 
\begin{tabular}{ p{2.5cm} p{1.2cm} p{1.2cm} p{1.2cm}} 
\hline  
Drying RH (\%) & 22 & 53 & 63   \\ \hhline{----}
$K_l$ (10$^{-21}$ m$^2$) & 2.2 & 1.5 & 1.9   \\ \hhline{----}
\end{tabular}\label{table:inverse_analysis}
\end{table}

\subsection{Influence of vapour diffusion } 
\begin{figure}[!ht] 
	\centering 
	\subfigure[Drying at RH=22\%.]{
		\includegraphics[width=6cm]{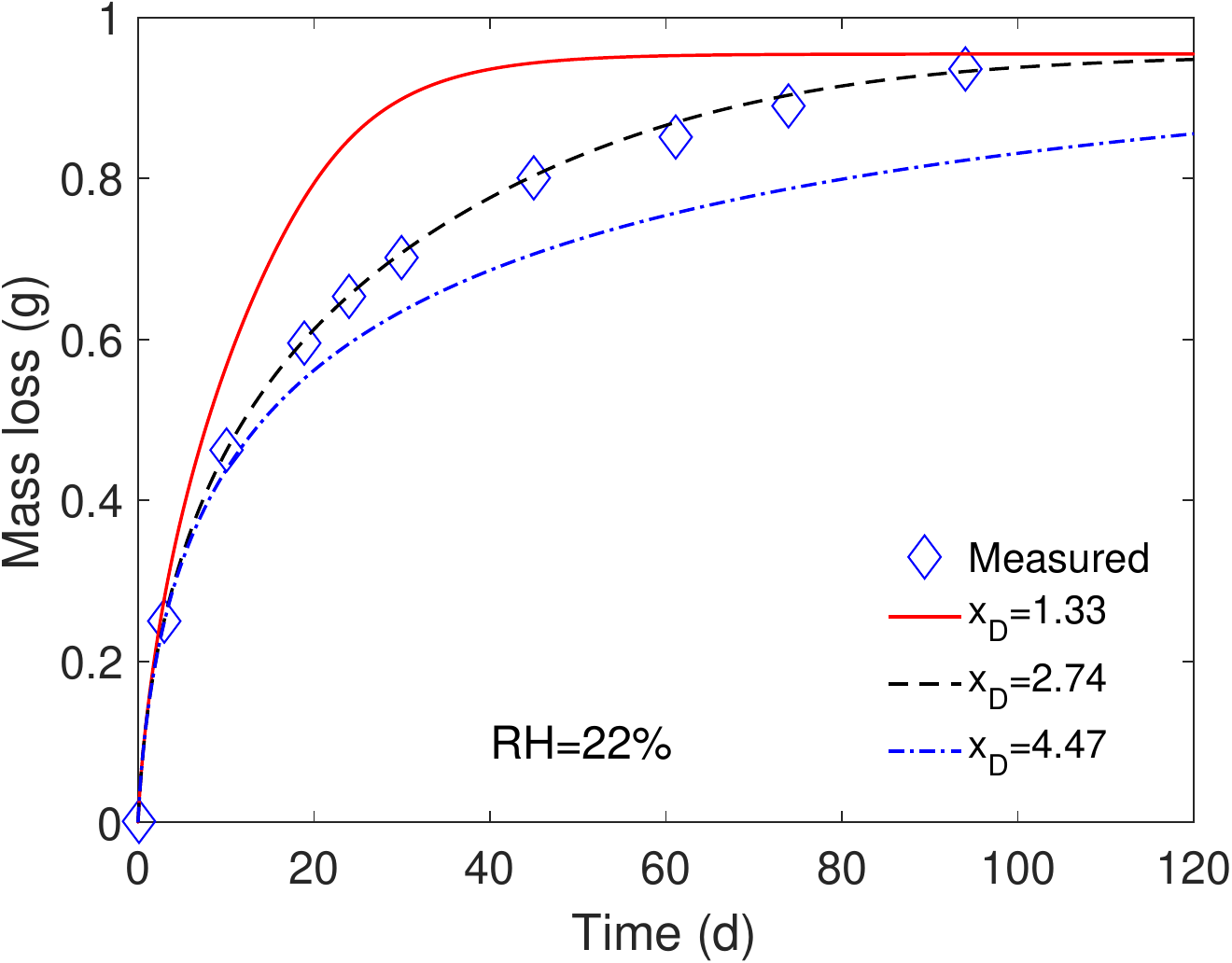}
		}\\
	\subfigure[Drying at RH=53\%.]{
		\includegraphics[width=6cm]{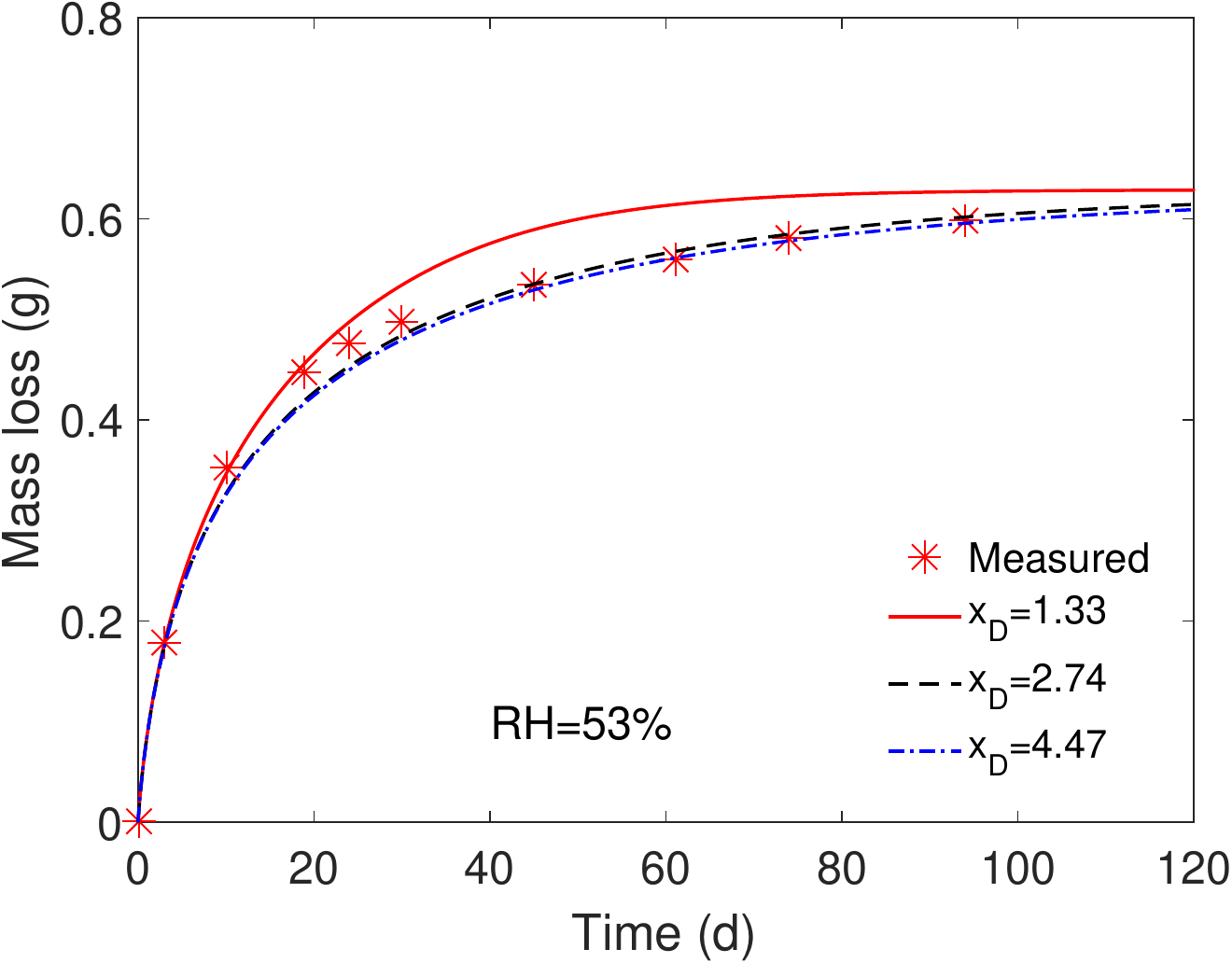}
		}\\
	\subfigure[Drying at RH=63\%.]{
		\includegraphics[width=6cm]{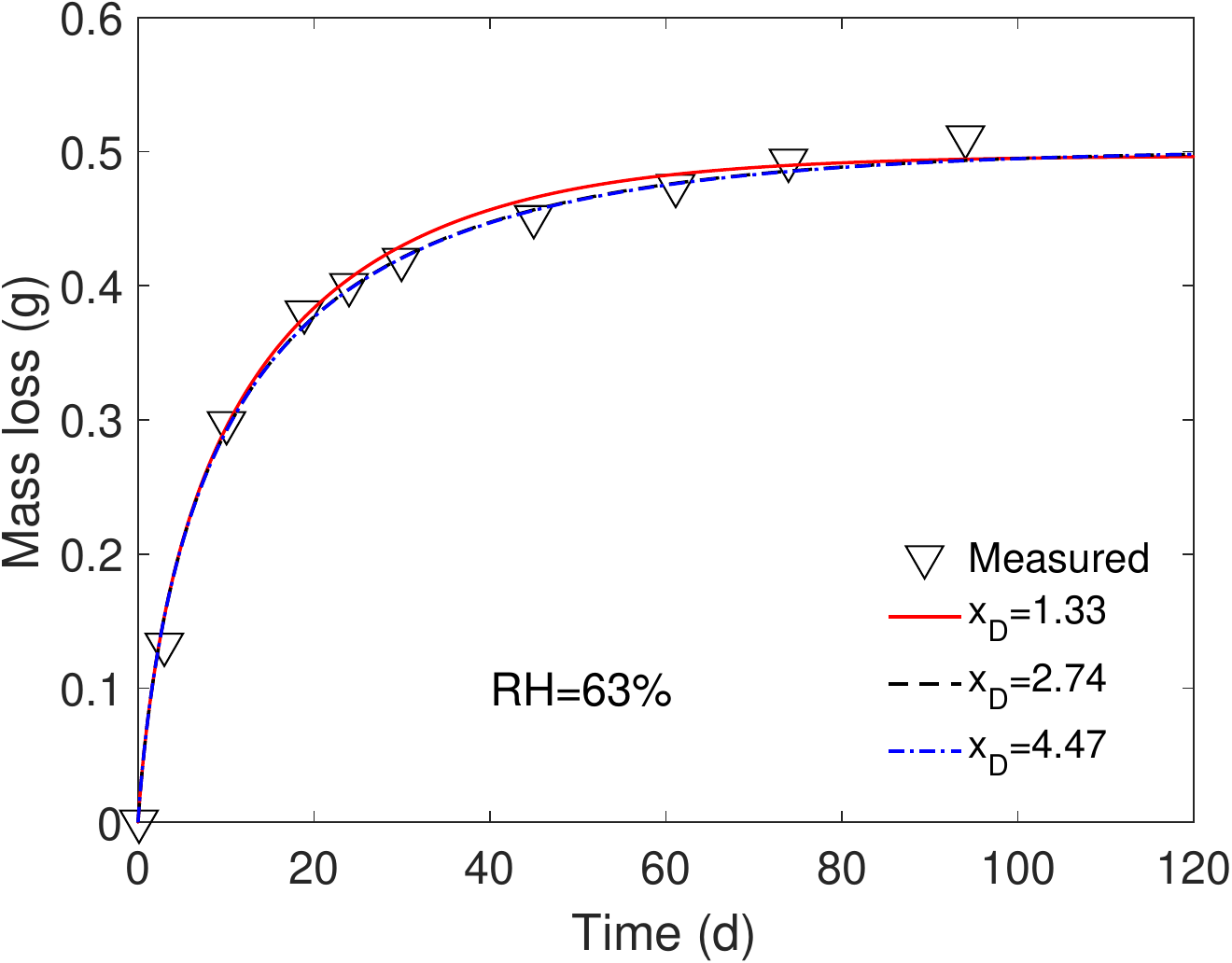}
		}
\caption{Comparison of the effect of vapor diffusion on simulated mass loss. The case of $x_D = 4.47 $ is used in this study to inversely determine $K_l$. }
\label{fig:vapor_diffusion} 
\end{figure}

Theoretically, $ K_l $ should be independent of the external drying conditions. The previous study~\cite{Zhang2016} argued that the discrepancy between $K_l$ determined at different RHs was thought to be affected by the preparation of different specimens of the same mix design. Nonetheless, the batch effect is minimized in this study. One of possible reasons is that the effect of vapor diffusion varies with the drying condition, because it was found that vapor diffusion gradually becomes more important with the decrease of external RH~\cite{Zhang2017DT}. 
In the above analysis, we assumed that $x_D=2.74$ in Eq.~\eqref{eq:resistance_factor} is suitable for all cases. %, while this may be not valid for all drying cases. 
To evaluate this assumption, we take different values for $x_D$ (the main parameter controlling the rate of vapor diffusion) to simulate the drying mass loss curve: $x_D = 1.33 $ was proposed by Millington and Quirk~\cite{Millington1961}, $x_D = 2.74 $ is suggested by Thiery et al.~\cite{Thiery2007}, and  $x_D = 4.47 $ is the value that we found for cementitious materials~\cite{Zhang2016} based on calibration by the measured $D_{eff}$ curve~\cite{VBB2007b}. Results of three drying conditions are compared in Fig.~\ref{fig:vapor_diffusion}. It is clear that the influence of $x_D$ is closely related to the drying condition, as it is more significant for drying at low RHs. The influence decreases with the increase of external RH and eventually at RH = 63\% the influence vanishes.

It was reported that $x_D$ varies with materials~\cite{Zhang2016} and a universal value that can be used for all cementitious materials could not be found. Since there is no generally accepted value for $x_D$, it is a good practice to choose the drying condition that has the negligible effect of vapor diffusion on the total mass loss. 
The goal here is to understand the range of RH under which the influence of vapor diffusion in the moisture transport model is negligible. Actually, this can be assessed by calculating the effective diffusion coefficient considering the moisture transport as a pure diffusion-like process. Based on Eqs.~\eqref{eq:mass_balance}, \eqref{eq:liquid_tansport} and \eqref{eq:vapor_diffusion}, the effective diffusion coefficient $D_{eff}$ is written as 

\begin{eqnarray}
\label{eq:Deff} 
D_{eff} =  k_{rl} \dfrac{K_l}{\phi\eta_l} \dfrac{\mathrm{d} P_c }{\mathrm{d} S} + \left( \dfrac{M_v}{\rho_l RT}\right)^2 D_{v0} f(S,\phi) \dfrac{P_{vs} RH}{\phi} \dfrac{\mathrm{d} P_c }{\mathrm{d} S} 
\end{eqnarray}

\begin{figure}[!ht] 
	\centering 
		\includegraphics[width=6cm]{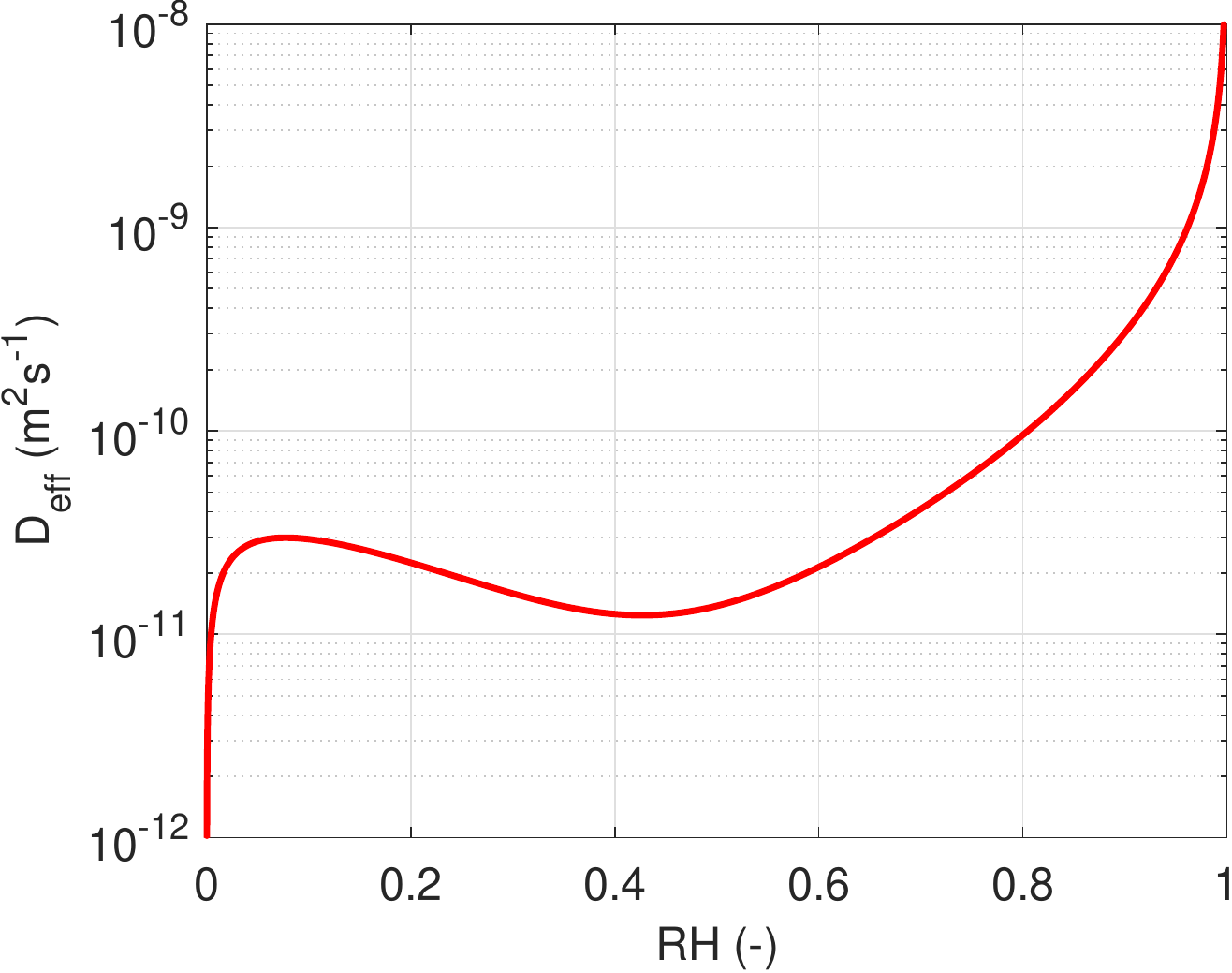}
\caption{Calculated $D_{eff}$ curve ($K_l = 1.9 \times 10^{-21}$ m$^{2}$ and $x_D = 2.74$). Note that $D_{eff}$ is material-dependent and thus for other materials the shape of the curve should be different to the one in this figure.}
\label{fig:Deff} 
\end{figure}

The plot of $D_{eff}$ vs. RH is shown in Fig.~\ref{fig:Deff} which displays a bumpy curve with one peak at low RH and one at RH close to 1. The lowest point between them is found at RH $\approx$ 43\%, which is considered as the demarcation between the liquid and vapor transport. Below this point, vapor diffusion dominates the moisture transport; above this point, the advection of liquid becomes significant. Clearly, RH = 53\% and 63\% are located in the liquid water dominant region. However, RH = 53\% is very close to the demarcation point and liquid transport is not considerably higher than vapor diffusion, so that we can still see the influence of $x_D$ on the calculated mass loss curve (see Fig.~\ref{fig:vapor_diffusion}b).
From this point of view, we are able to say that $K_l$ determined by the drying test at RH=63\% is more accurate than values obtained by drying at lower RHs. 
Therefore, our recommendation is that if one wants to use the inverse analysis method to determine $K_l$, it is better to avoid using measured mass loss curves at low RHs (i.e., below the demarcation point on the effective diffusion coefficient curve). 

Another implication from this discussion is that an appropriate model that targets on the whole range of RH must include the contribution from both liquid and vapor. Some models in the literature that consider the effective diffusion coefficient/diffusivity as a monotonically increasing function with water content/degree of saturation are not suitable for moisture transport at low RHs, because they basically ignore the transport of water vapor. 

\subsection{Overestimation of KC and KT equations} 
In agreement with most literature, this study finds that the KC and KT equations overestimate $K_l$ of cement pastes. 
Halamickova et al.~\cite{Halamickova1995} pointed out that the KT relations may work better for systems with highly interconnected capillary pore networks than for systems where the fine nanostructure of gel pores dominates the transport, as is the case in most cementitious materials. This can help to explain why the permeability calculated by the KT equations is larger. Indeed, as originally explained by Katz and Thompson, their theory is valid for materials with large pores and mono-modal PSD centered in the capillary range~\cite{Johnson1986}. %, in which the surface effects can be neglected compared to the pore size~\cite{Johnson1986}. 

In addition to the inherent reasons, there are some deficiencies in the measured data for KC and KT equations: 
\begin{itemize}
\item[] 1) The data used in these equations are measured by MIP, which is not sensitive to the small pores. %The structural difference of cementitious materials is one reason. 
Pores in hydrated materials can be generally divided into capillary pores and gel pores. Capillary pores are easily detected by MIP, but not gel pores, whose contribution to the permeability of cementitious materials is non-negligible~\cite{Scherer2007a}. The contribution of small pores to the PSD is not fully included in MIP data and therefore measured surface area is much lower than other techniques, so that it leads to a higher $K_l$ in the KC equation. 
%moisture transport in cement paste is not just controlled by capillary pore but also gel pores. 
\item[] 2) Another basic problem with MIP is that the specimens are dried before measurements. As pointed out in the literature, the microstructure of the material is altered whatever the selected method for drying (oven-drying, vacuum drying, freezing drying or solvent exchange)~\cite{Zhang2011}. 
The different measured PSD curves for two drying methods in the present study also show that the pretreatment conditions have an influence on the measured PSD. 

%However, the difference of the critical pore diameter between two drying methods is only about 10 nm and much smaller than one order of magnitude. IPA exchange drying is considered to be most useful to preserve the microstructure. 
%
%The main reason is that in a MIP measurement the specimen must be totally dried.  Supercritical drying can be considered as a better drying method to preserve the microstructure~\cite{Zhang2011,Zhang2017SCD}. %while no detailed results have still appeared in the literature.
\item[] 3) The choice of the critical diameter also contributes the difference. The critical pore diameter was defined as the pore size above which a connected path can form from one side of the specimen to the opposite side. It is reflected by the inflection point of the cumulative mercury intrusion cure. The determination of the critical diameter holds an accuracy of $ \pm 15\% $~\cite{Ma2014mercury}. 
According to the definition of the critical diameter, it is most likely the characteristic pore diameter of gel pores $d_{gp}$, which is much smaller than $d_{c}$~\cite{Aligizaki2006pore}. If $ d_{gp} $ is used in Eqs.~\eqref{eq:KTI} and \eqref{eq:KC_2}, the predicted permeability should be about one or two orders of magnitudes lower, and much closer to the values from the BB and inverse analysis methods. However, the choice is not arbitrary and is controlled by the critical porosity $\phi_c$ which varies with materials. 
\end{itemize}

Hence, if the measured microstructure is more representative of the original one, it is expected to yield more accurate $K_l$. Experimental methods that can measure the PSD of the saturated specimen are suggested for the future studies, such as nuclear magnetic resonance (NMR) and thermoporometry. 

%the precision of the normalized conductivity and critical pore radius values creates variability of the predicted permeability results but not to as large as an extent. 

%\subsection{Limitations of moisture transport model} 
%the method to choose the equation for relative permeability 
%
%For this kind of model, the microstructural change is not directly considered by any equation implemented in the model. However, the input data - the measured sorption isotherm - actually include the information.  
%
%this model was verified by both drying kinetics and measured water saturation profiles. 

%\subsection{Improving sorptivity method}
%We have attempted to fit the measured sorptivity vs. RH with $K_l$ and $n_i$ as free parameters. This presumes that $K_l$ and $n_i$ are the same for all sorptivity tests, regardless of the initial conditions. However, this method did not work because $\tau_D$ and $K_l$ are combined together in Eqs.~\eqref{eq:intrinsic1} and~\eqref{eq:Zhou_solution_Dl0} (with $\dfrac{K_l}{\tau_D}$). $\tau_D$ decreases with the increase of $n_i$. Hence, there are numerous combinations of $n_i$ and $K_l$ that can provide good fitting. 

\subsection{Permeability measured by other fluids} 
% permeability measured by water, alcohols (methanol, isopropanol) and gas (oxygen and nitrogen) are different 
%permeability measured by gas and solvent (they all change the natural microstructure of cementitious materials)

% gas permeability measurement needs dried specimen it may induce cracks 
%a dried specimen has different microstructure to the water saturated one. 
%The presence of drying cracks was found to have little effect on the rate of capillary absorption of dried mortar (Wu et al. 2015). Similarly, Bisschop and van Mier (2008) found little influence of microcracks on the rate of drying, although the drying rate decreased as the volume fraction of impermeable aggregate increased. Bazant et al. (1987) found that the permeability and diffusivity were raised by cracks, but much less than expected on the basis of the crack widths observed at the surface. They concluded that the cracks must be interrupted by constrictions that inhibit transport. Nevertheless, microstructural modification at the pore scale certainly occurs during drying, because the permeability of a dried and resaturated cement paste is more than an order of magnitude higher than in the undried body (Powers et al. 1955; Hearn 1996).

% measurements with alcohols change the microstructure of materials 
The permeability is often measured by other gases and liquids in the literature. 
For gas permeability measurements, specimens must be dried or partially dried. The problem is that the morphology of C-S-H can be significantly changed by drying. By observing under the environmental scanning electron microscope (ESEM), Fonseca and Jennings~\cite{Fonseca2010} reported that a slow drying, such as drying at high RH for several days, creates a bumpy morphology around the cement grain, which is very different to the needle-like structure in the rapidly dried specimens. Based on the comparison of different drying methods, Zhang and Scherer~\cite{Zhang2017SCD} and Zhang et al.~\cite{Zhang2017SEM} found that no matter which drying method is used, as long as water is removed from the pore network, the morphology of hydration products is altered and thin sheets between fine fibers are created. These effects, either forming bumpy surface or creating thin sheets, will increase the complexity of the microstructure. Together with coarsening the pore size distribution (e.g.,~\cite{Zhou2017}) and inducing micro-cracks (e.g.,~\cite{De2008analysis}), the transport properties would be affected. Therefore, we should not directly compare permeabilities measured by water with that by gases, because the microstructure of the specimens is different to each other. 

Loosveldt et al.~\cite{Loosveldt2002} found that water permeability was systematically lower (by 1 - 2 orders of magnitudes) than that of argon gas, and ethanol permeability was between them. A similar conclusion was reported by Zhou et al.~\cite{Zhou2016indirect} that ethanol permeability is about 2 - 3 orders of magnitudes higher than water permeability and close to the gas permeability with the correction of Klinkenberg slippage effect. However, these results are questionable. First, ethanol has been proved to react with hydration products (essentially calcium hydroxide)~\cite{Zhang2011}. Second, ethanol permeability was measured after gas permeability measurements on the same specimen. As mentioned above, the microstructure of dried specimen has been altered compared to the original water-saturated one. A more acceptable way is to do solvent exchange with the water-saturated specimen before measuring the solvent permeability. 
By using this procedure, Hearn~\cite{Hearn1996} reported that there is only slight difference between water and IPA permeabilities. The difference is due to water having stronger chemical interactions than IPA with the hydration products. Nevertheless, recent studies by Zhou et al.~\cite{Zhou2017} reported that IPA permeability is about 2 - 3 orders of magnitudes higher than water permeability. They found that the peaks of PSD measured by NMR shift to high pore size range for IPA treated specimens, meaning that the microstructure is coarsened by IPA replacement. It is hard to believe that IPA replacement causes permeability 2 - 3 orders of magnitudes higher than the one measured with water saturated specimen since IPA replacement was shown as a preferable drying method to preserve the microstructure~\cite{Konecny1993,Beaudoin1987,Zhang2011,Zhang2017SEM}. The most possible reason is that some microcracks are induced during solvent exchange as reported in our previous study~\cite{Zhang2017SCD} that the damage is more serious for a large specimen ($\geq$ 8 mm in diameter). The thickness of the specimen in~\cite{Zhou2017} is 20 – 25 mm, so damages most likely happened during exchanging with IPA. Another fact that can lead to damage during IPA replacement is the exchange duration. Vichit-Vadakan and Scherer~\cite{Vichit2002} reported that after about a 6-week exchange the measured porosity of cement paste dramatically increased to the same value as the oven dried specimen at 105 $^\circ$C, and they suspected damages because of the long exchange duration. 

%
%\subsection{Sorption isotherms measurements}
%gradually drying on one specimen 
%
%drying different specimens at different RHs 
%
%gradual drying may significantly alter the microstructure and further changes the transport properties, such as creating more small or isolated pores and reducing permeability. 

%%%%%%%%%%%%%%%%%%%%
% NEW SECTION
%%%%%%%%%%%%%%%%%%%%
\section{Conclusion }
To eliminate the artifacts (e.g., types of cements, specimen preparation and curing conditions) in the measurement of water permeability, this study prepared all specimens from a long cylinder for a variety of permeability determination methods. specimens from the same cylinder were also used for experiments to obtain the input and calibration data for a moisture transport model. 
By comparing the simulated and measured mass loss curves, we found that the KC, KTII and sorptivity methods yield significantly greater values of $K_l$ which lead to overestimation of the mass loss. Permeability determined by the BB method provides a mass loss curve slightly lower than measured one, but it is the closest one among the compared methods.

The overestimation of the KC, KTII and sorptivity methods results from the fact that specimens used in these methods must be dried before the tests. As shown in our previous studies, any drying (either fully or partially) can significantly alter the microstructure of the material. In addition, MIP is not an appropriate method to detect the small pores in cementitious materials; thus, the measured surface area and critical pore size are far from the \enquote{true} ones in the water-saturated specimen. 
In addition, by comparing sorptivity measurements for different initial water content, we found that the use of specimens with high initial water content is inappropriate for the sorptivity method. 

The permeability values inversely determined by the moisture transport model based on the measured mass loss curves are very close to those measured by the beam-bending method. 
The most important notice put forward in this study for the use of the inverse analysis method is that drying tests must be done at high RHs (63\% in this study) to reduce the effect of vapor diffusion on the determination of water permeability. 
% If one has to use drying at low RHs, one should be aware of that the choice of value for $x_D$ may affect the result.  

This study has tried to minimize the batch effect, but permeabilites determined by a variety of methods are still much different to each other. This may indicate that the effect of choosing the water permeability determination method is more important than eliminating the batch effect. 
 
% NEW section 
\section*{Acknowledgment}
ZZ would like to thank Professor Chunsheng Zhou from HIT for discussion of using sorptivity method to determine permeability. 
%\thispagestyle{empty}

%\clearpage  
%\usepackage[refpage]{nomencl} 
%\renewcommand{\nomname}{List of Notations} 
%\renewcommand*{\pagedeclaration}[1]{\unskip\dotfill\hyperpage{#1}} 
%\makenomenclature 
%\usepackage{makeidx} 
%\makeindex 
%\appendix 

%\printindex

%\printnomenclature 
%\printglossary 


%\section*{References}
\bibliographystyle{plain}
%\bibliography{References}

\end{document}